\documentclass[
    preprintnumbers,
    twocolumn,
    showkeys,
    superscriptaddress,
    showpacs,
    nofootinbib,
    longbibliography,
    pra,
    notitlepage
]{revtex4-1}

\usepackage[utf8]{inputenc}
\usepackage{mathtools}
\usepackage{physics}
\usepackage{csquotes}
\usepackage{amsmath}
\usepackage{amsthm}
\usepackage{amsfonts}
\usepackage[caption=false]{subfig}
\usepackage{graphicx}
\usepackage[section]{placeins}
\usepackage{setspace}
\usepackage{algorithm}
\usepackage{algpseudocode}
\usepackage{verbatim}
\usepackage{bbm}
\usepackage{slashed}
\usepackage[colorlinks=true,linkcolor=blue,citecolor=blue,urlcolor=blue]{hyperref}

\newtheorem{statement}{Statement}

\begin{document}

\title{
Long-time fermionic quantum transport with controlled full-state error
using an adaptive reservoir-mode window
}

\author{Mikhail Umanskii}
\email[]{mix.umanskiy@gmail.com}
\affiliation{Russian Quantum Center, Bolshoy Bulvar 30, bld. 1, Skolkovo, Moscow 121205, Russia}
\affiliation{Moscow Institute of Physics and Technology, Institutsky lane 9, Dolgoprudny, Moscow region, 141700, Russia}
\author{Nataliya Arefyeva}
\email[]{arefnat8@gmail.com}
\affiliation{Russian Quantum Center, Bolshoy Bulvar 30, bld. 1, Skolkovo, Moscow 121205, Russia}
\affiliation{Faculty of Physics, M.V.Lomonosov Moscow State University, Leninskie Gory, Moscow 119991, Russia}
\author{Georgy Sultanov}
\email[]{gfaberlane@gmail.com}
\affiliation{Russian Quantum Center, Bolshoy Bulvar 30, bld. 1, Skolkovo, Moscow 121205, Russia}
\affiliation{National Research Nuclear University MEPhI, Moscow, 115409, Russia}
\author{Alexey Rubtsov}
\email[]{ar@rqc.ru}
\affiliation{Russian Quantum Center, Bolshoy Bulvar 30, bld. 1, Skolkovo, Moscow 121205, Russia}
\affiliation{Faculty of Physics, M.V.Lomonosov Moscow State University, Leninskie Gory, Moscow 119991, Russia}
\author{Evgeny Polyakov}
\email[]{evgenii.poliakoff@gmail.com}
\affiliation{Russian Quantum Center, Bolshoy Bulvar 30, bld. 1, Skolkovo, Moscow 121205, Russia}

\begin{abstract}
Real-time simulations of interacting nanostructures coupled to fermionic reservoirs can require a growing number of environmental degrees of freedom to retain long-lived correlations. We introduce tape-recorder coarse graining, which reorganizes each noninteracting lead into incoming, active, and outgoing modes. The device is propagated with the active modes, while outgoing modes are stochastically sampled and removed once their remaining integrated coupling falls below a prescribed threshold. For each outgoing-mode truncation, we derive a nonperturbative upper bound on the infidelity between the exact and truncated full device-reservoir states over any prescribed finite interval. The bound depends on the mode's remaining coupling weight and finite-interval response factors. Numerically, the active-mode count saturates in time at fixed relative threshold and grows logarithmically as the threshold is reduced. We benchmark the method on a two-site quantum point contact at zero temperature and maximal bias. For Lorentzian reservoirs, the dynamics agrees with converged HEOM calculations and the steady-state current with the Landauer-Büttiker result. For flat-band reservoirs with algebraically decaying correlations, it agrees with direct Schrödinger evolution before finite-size recurrences and reproduces the Landauer-Büttiker stationary current, while finite exponential HEOM decompositions remain unconverged. For interacting contacts, the method yields Coulomb-blockade peak splitting. In the noninteracting driven limit, it agrees with an exact Floquet Green-function calculation and reproduces coherent current suppression under periodic driving, which persists at finite Coulomb repulsion. Together, these benchmarks show that tape-recorder coarse graining enables practical long-time simulations of the full device-reservoir state in interacting fermionic transport.
\end{abstract}

\maketitle

\section{Introduction}
Quantum transport through quantum dots, molecular junctions, and quantum point contacts is a central nonequilibrium many-body problem \cite{Kouwenhoven1997,Nitzan2003,PlateroAguado2004,Kohler2005,Ryndyk2016,Landi2022}. Stationary noninteracting transport is efficiently described by the Landauer-Büttiker formalism \cite{Landauer1957,Buttiker1986}, whereas real-time simulations of interacting, driven, or strongly non-Markovian devices require explicit treatment of nonequilibrium device–reservoir dynamics \cite{Jauho1994,BreuerPetruccione2007,deVegaAlonso2017}. The main computational challenge is the continuous growth of entanglement between the finite device and its macroscopic reservoirs \cite{Wolf2014,RamsZwolak2020,Kohn2021,Lerose2021,He2017}.

Controlled many-body methods often rely on regime-specific small parameters or compression mechanisms. Perturbative diagrammatic expansions rely on weak interactions, renormalization-group methods on scale separation, Markovian master equations on short bath memory, and tensor-network methods on compressible entanglement \cite{NegeleOrland1988,BreuerPetruccione2007,Bulla2008,AltlandSimons2010,Schollwoeck2011}. Interacting, driven, and non-Markovian transport can fall outside these favorable regimes. Systematically convergent approaches such as hierarchical equations of motion (HEOM) \cite{Jin2008,TanimuraKubo1989,IshizakiFleming2009}, time-evolving matrix product operators (TEMPO) \cite{Strathearn2018,Chen2024}, and chain-mapped tensor networks can capture non-Markovian dynamics \cite{Chin2010,deVega2015,Tamascelli2018}. Their convergence is controlled by method-specific truncations of hierarchy depth, memory length, chain length, or bond dimension \cite{Jin2008,Strathearn2018,deVega2015,Park2024,Ng2023}. Maintaining a prescribed accuracy at long times can require increasing computational resources. In demanding regimes, simulations are therefore often restricted to short or intermediate times, complicating access to steady states and long-time driven responses \cite{Ng2023,RamsZwolak2020,Cohen2015}.

We instead characterize each reservoir mode by its coupling weight, the time-integrated intensity of its coupling to the device. Modes whose coupling weight falls below a prescribed fraction of the largest weight are omitted from the active dynamics \cite{Polyakov2022RealTime,Polyakov2022Bandlimited,ArefyevaPolyakov2026}. For each outgoing-mode truncation, we derive a nonperturbative bound on the resulting full-state infidelity over any prescribed finite interval. The criterion does not require weak instantaneous coupling. The retained modes define a moving active subspace. We show numerically that its dimension saturates in time at a fixed threshold and grows approximately logarithmically as the threshold is tightened.

We construct this moving subspace through time-dependent rotations of the reservoir modes. In this basis, each reservoir mode follows the same lifecycle. Before entering the active window, its accumulated coupling weight is negligible, and the mode is omitted from the dynamics. As the mode enters the active window, its accumulated coupling weight exceeds the prescribed threshold, and it is included in the dynamics. It becomes outgoing once its remaining integrated coupling weight falls below the corresponding threshold. We call these stages incoming, active, and outgoing \cite{Polyakov2022RealTime,Polyakov2022Bandlimited}. A reservoir mode passes through them like a segment of tape moving past a recording head, motivating the term ``tape-recorder'' coarse graining~\cite{ArefyevaPolyakov2026}.

An outgoing mode carries a record of the preceding evolution, while its remaining coupling over the prescribed simulation interval lies below the truncation threshold \cite{ArefyevaPolyakov2026}. The full-state error caused by neglecting this residual interaction is bounded. Stochastic sampling of this occupation represents the partial trace over the outgoing mode while allowing the device and active reservoir modes to continue evolving as a wavefunction \cite{WisemanMilburn1993,Daley2014,Polyakov2022RealTime}.

The tape-recorder construction was introduced for a local quench in a noninteracting integrable environment and developed for bosonic examples \cite{Polyakov2022RealTime,Polyakov2022Bandlimited,ArefyevaPolyakov2026}. Here we extend it to fermionic reservoirs and to transport through a contact between two leads. The fermionic extension raises two technical issues. First, fermionic anticommutation affects the ordering of the system and reservoir operators \cite{NegeleOrland1988,AltlandSimons2010}. The annihilation and creation source channels combine through the canonical anticommutation relations, so the single-event error bound depends on the remaining coupling weight and the response of the retained dynamics without an assumption on the outgoing-mode occupation. Second, transport reservoirs generally start from finite-density Gaussian states rather than vacuum states \cite{Datta1995}. In the zero-temperature, fully biased setup considered here, the left lead is filled and the right lead is empty across the full lead band. A particle-hole transformation maps the filled lead to a quasiparticle vacuum \cite{TakahashiUmezawa1975,Schwarz2018}, while preserving the mode-space light-cone structure. The reservoir-mode construction is then applied to the quasiparticle fields.

As a proof of concept, we apply the method to a two-site interacting quantum point contact coupled to semi-infinite tight-binding leads. For the bounded lead spectral densities considered here, the maximum time-integrated coupling weight remains finite as the simulation interval is extended. Numerically, the active-mode count saturates at a fixed threshold. This saturation keeps the peak Hilbert-space dimension of each stochastic trajectory bounded over the simulated time window. The online propagation cost per trajectory therefore grows linearly with the number of time steps.

We validate the method against Landauer-Büttiker theory, exact Floquet Green-function results, and direct Schrödinger evolution. For Lorentzian reservoirs, it agrees with converged HEOM calculations, whereas the finite exponential HEOM decompositions tested for a flat band remain unconverged. For interacting contacts, the current develops the two-peak structure associated with Coulomb blockade \cite{Kouwenhoven1997}. In the noninteracting driven limit, the method reproduces coherent current suppression in agreement with an exact Floquet Green-function calculation \cite{Kohler2005}. The suppression persists at finite Coulomb repulsion with interaction-dependent corrections. These driven results connect to recent studies of nonequilibrium transport and Floquet thermalization breakdown in fermionic chains \cite{Gamayun2021,Dudinets2025}.

Section \ref{sec:model} specifies the QPC model, quench protocol, and particle-hole mapping. Section \ref{sec:tape_recorder} presents the tape-recorder construction. Section \ref{sec:results} presents the numerical benchmarks and the interacting and driven applications. Section \ref{sec:discussion} compares the method with existing approaches and outlines extensions to interacting reservoirs and multiterminal geometries. Section \ref{sec:conclusion} summarizes the main results.

\section{Model}\label{sec:model}

\subsection{Hamiltonian and lead structure}
We consider a two-site interacting quantum point contact coupled to two semi-infinite noninteracting fermionic leads. The total Hamiltonian is
\begin{equation}
\hat{H}(t) = \hat{H}_S(t) + \hat{H}_L + \hat{H}_R + \hat{H}_{\text{coup}}.
\end{equation}
The leads are spinless tight-binding chains with nearest-neighbor hopping amplitude $t>0$:
\begin{equation}
\hat{H}_\alpha = -t \sum_{j=0}^{\infty} \left( \hat{c}_{\alpha,j+1}^\dagger \hat{c}_{\alpha,j} + \text{h.c.} \right), \quad \alpha \in \{L,R\}.
\end{equation}

The operators $\hat c_{\alpha,j}$ and $\hat c_{\alpha,j}^\dagger$ annihilate and create a fermion on site $j$ of lead $\alpha$ and obey the canonical anticommutation relations $\{\hat c_{\alpha,i},\hat c_{\beta,j}^\dagger\}=\delta_{\alpha\beta}\delta_{ij}$. The dispersion $\varepsilon_\alpha(k)=-2t\cos k$ has bandwidth $W=4t$.

The QPC consists of two fermionic sites with gate-tunable on-site energy $\varepsilon_d(t)$ and intersite Coulomb repulsion $U$:
\begin{equation}
\hat{H}_S(t) = \varepsilon_d(t) (\hat n_0 + \hat n_1) + U \hat n_0 \hat n_1 - t \left(\hat d_0^\dagger \hat d_1 + \hat d_1^\dagger \hat d_0\right).
\end{equation}
Here $\hat d_m$ annihilates a fermion on QPC site $m\in\{0,1\}$, and $\hat n_m=\hat d_m^\dagger\hat d_m$. The lead coupling is local and symmetric:
\begin{equation}\label{eq:coupling}
\hat H_{\text{coup}} = -t\left( \hat c_{L,0}^\dagger \hat d_0 + \hat d_0^\dagger \hat c_{L,0} \right) -t\left( \hat c_{R,0}^\dagger \hat d_1 + \hat d_1^\dagger \hat c_{R,0} \right).
\end{equation}

We use the same hopping amplitude $t$ within the QPC, in the leads, and across both QPC-lead contacts. This choice probes a moderate-to-strong-coupling regime with pronounced non-Markovian memory.

\subsection{Quench protocol and initial state}
The transport protocol begins with a sudden QPC-lead coupling quench at $t=0$. For $t<0$, $\hat H_{\mathrm{coup}}=0$, and the leads and QPC are prepared in the product state:
\[
|\Psi(0)\rangle = |\Psi_L\rangle \otimes |\Psi_R\rangle \otimes |\Psi_{\mathrm{QPC}}\rangle .
\]
The left lead is fully occupied at zero temperature, with chemical potential $\mu_L=+2t$ at the upper band edge:
\[
|\Psi_L\rangle = \prod_{j=0}^{\infty} \hat c_{L,j}^{\dagger}|0\rangle .
\]
The right lead is empty, $|\Psi_R\rangle=|0\rangle$, with $\mu_R=-2t$ at the lower band edge. The resulting bias window $\Delta\mu=4t$ spans the full lead bandwidth. The QPC is initially in the empty state $|\Psi_{\mathrm{QPC}}\rangle=|00\rangle$.

The local quench launches excitations that propagate ballistically into the leads. The tight-binding dispersion gives a maximum group velocity $v_{\max}=2t$, consistent with a Lieb-Robinson light cone \cite{Lieb1972}. At any finite time, the dynamically relevant lead region is therefore finite up to exponentially small tails.

\subsection{Particle-hole mapping to a vacuum reference}
The reservoir-mode construction is formulated relative to a quasiparticle vacuum. Since the left lead is fully occupied, we introduce hole operators through the particle-hole transformation:
\[
\hat{\tilde{c}}_{L,j} = \hat c_{L,j}^\dagger, \quad \hat{\tilde{c}}_{L,j}^\dagger = \hat c_{L,j}.
\]
We leave the right-lead operators unchanged, $\hat{\tilde c}_{R,j}=\hat c_{R,j}$. The filled left lead is thereby mapped to a hole vacuum, while the empty right lead remains a particle vacuum. Thus both reservoirs are empty in the quasiparticle representation. Their combined initial state is the quasiparticle vacuum $|\tilde 0\rangle_{\mathrm{res}}$, satisfying:
\[
\hat{\tilde c}_{\alpha,j} |\tilde 0\rangle_{\mathrm{res}} = 0, \quad \alpha\in\{L,R\}.
\]
The transformed operators obey the canonical anticommutation relations:
\[
\{\hat{\tilde c}_{\alpha,i}, \hat{\tilde c}_{\beta,j}^{\dagger} \} = \delta_{\alpha\beta}\delta_{ij}.
\]

The left-lead Hamiltonian becomes:
\begin{equation}
\hat H_L \xrightarrow{\mathrm{p-h}} t\sum_{j=0}^{\infty} \left( \hat{\tilde c}_{L,j+1}^{\dagger}\hat{\tilde c}_{L,j} + \text{h.c.} \right).
\end{equation}
The quasiparticle dispersion is therefore $\tilde{\varepsilon}_L(k)=+2t\cos k=-\varepsilon_L(k)$. For the real nearest-neighbor chain, this sign reversal complex conjugates the single-particle propagator and therefore leaves the spatial probability profile, bandwidth, and maximum speed $v_{\max}=2t$ unchanged. Physically, tunneling from the filled left lead creates a hole, whereas tunneling into the empty right lead creates a particle. In the quasiparticle representation, both are excitations above a vacuum and propagate away from the QPC into their respective leads, while representing the same left-to-right physical particle current. To express the QPC in the same quasiparticle representation, we define $\hat{\tilde d}_0=\hat d_0^\dagger$ and leave the second-site operator unchanged, $\hat{\tilde d}_1=\hat d_1$. Hence $\hat n_0=1-\hat{\tilde n}_0$ and $\hat n_1=\hat{\tilde n}_1$. Up to the additive scalar $\varepsilon_d(t)$, the QPC Hamiltonian becomes:
\begin{equation}
\begin{gathered}
\hat{\tilde H}_S(t) = 
-\varepsilon_d(t)\hat{\tilde n}_0 + [\varepsilon_d(t)+U]\hat{\tilde n}_1 - 
\\ -
U\hat{\tilde n}_0\hat{\tilde n}_1 - t\left( \hat{\tilde d}_0\hat{\tilde d}_1 + \hat{\tilde d}_1^\dagger\hat{\tilde d}_0^\dagger \right). 
\end{gathered}
\end{equation}
The physical empty QPC state corresponds to $\hat{\tilde n}_0=1$ and $\hat{\tilde n}_1=0$; it is therefore not part of the reservoir quasiparticle vacuum.

\subsection{Open-system structure and transport observables}
For the remainder of the paper, $\hat H(t)$ denotes the Hamiltonian in the quasiparticle representation. It has the standard open-system form:
\begin{equation}
\hat H(t) = \hat{\tilde H}_S(t) + \sum_{\alpha=L,R}\hat H_\alpha + \sum_{\alpha=L,R} \left( \hat V_\alpha^\dagger\hat a_{\alpha,0} + \hat a_{\alpha,0}^\dagger\hat V_\alpha \right).
\end{equation}
Here $\hat H_L$ is the band-inverted left-lead Hamiltonian, whereas $\hat H_R$ is unchanged. The coupling-site operators are
\[
\hat a_{L,0}=\hat{\tilde c}_{L,0},
\quad
\hat a_{R,0}=\hat{\tilde c}_{R,0}=\hat c_{R,0},
\]
and the system-side coupling operators are
\[
\hat V_L=t\hat{\tilde d}_0,
\quad
\hat V_R=-t\hat{\tilde d}_1.
\]
The primary physical observables are the QPC occupations $\langle\hat n_m(t)\rangle$ and the lead currents. We orient both currents from left to right and define
\begin{align}
I_L(t)
&= -\frac{d}{dt}\langle\hat N_L(t)\rangle,
&
I_R(t)
&= \frac{d}{dt}\langle\hat N_R(t)\rangle,
\end{align}
where $\hat N_\alpha=\sum_{j=0}^{\infty} \hat c_{\alpha,j}^{\dagger}\hat c_{\alpha,j}$.
For the right contact:
\begin{equation}
I_R(t) = i\,t \left\langle \hat c_{R,0}^{\dagger}(t)\hat d_1(t) - \hat d_1^{\dagger}(t)\hat c_{R,0}(t) \right\rangle .
\end{equation}
Particle-number conservation gives:
\begin{equation}
\frac{d}{dt} \left\langle \hat n_0(t)+\hat n_1(t) \right\rangle = I_L(t)-I_R(t).
\end{equation}
In a stationary state, the QPC occupation is time independent and hence $I_L=I_R$. In a periodic steady state, the currents need not be equal instantaneously, but their averages over one driving period coincide.

\section{Tape-recorder coarse graining}
\label{sec:tape_recorder}
Extended reservoirs are often represented by chains, but the part of a chain involved in the dynamics grows with time as excitations propagate away from the contact. Our aim is to retain as few reservoir degrees of freedom as possible while preserving a prescribed accuracy. We therefore employ the tape-recorder coarse-graining method introduced in Refs.~\cite{Polyakov2022RealTime,Polyakov2022Bandlimited} and formulated in terms of forward and backward light cones. Here we give a self-contained formulation specialized to fermionic transport.

The tape-recorder construction therefore uses a time-dependent reservoir basis in which the modes needed at a given time form a finite moving window, rather than truncating the leads in real space. This basis separates modes that have not yet interacted appreciably with the QPC, modes that remain relevant to its subsequent evolution, and modes whose remaining influence has become negligible.

\subsection{Active modes and basis choice}

Which lead modes must be retained at time $t$? A mode can be excluded from the propagated state for either of two physical reasons. It may not yet have interacted appreciably with the QPC, or it may have already interacted but have only negligible influence left over the remaining interval $[t,T]$. The active modes are those for which neither exclusion applies: they have already coupled appreciably to the QPC and can still affect its subsequent evolution.

We first characterize whether a mode has already coupled. The QPC couples to lead $\alpha$ through the contact operator $\hat a_{\alpha,0}$. For a normalized fermionic lead mode $\hat\kappa_\alpha^\dagger$, the anticommutator
\begin{equation}
\chi_\alpha[\kappa;t] = \left\{\hat a_{\alpha,0}(t),\hat\kappa_\alpha^\dagger\right\}
\label{eq:mode_coupling_amplitude}
\end{equation}
gives its instantaneous coupling amplitude to the QPC. The coupling accumulated over the elapsed interval $[0,t]$ is quantified by
\begin{equation}
\mathcal I_\alpha^+[\kappa;t] = \int_0^t d\tau\, \left|\chi_\alpha[\kappa;\tau]\right|^2.
\label{eq:forward_weight}
\end{equation}
This elapsed coupling weight measures the extent to which the mode has already participated in the QPC dynamics. While $\mathcal I_\alpha^+$ remains small, the mode can be left out of the propagated state. Once it becomes appreciable on the chosen accuracy scale, the mode has arrived.

The second criterion determines when an arrived mode no longer needs to be retained. We quantify the coupling over remaining interval $[t,T]$ by
\begin{equation}
\mathcal I_\alpha^-[\kappa;t,T] = \int_t^T d\tau\, \left| \chi_\alpha[\kappa;\tau] \right|^2.
\label{eq:backward_weight}
\end{equation}
If $\mathcal I_\alpha^-$ becomes negligible after the mode has arrived, its residual influence over the prescribed simulation interval is negligible and the mode can be removed from the propagated state. Thus, $\mathcal I_\alpha^+$ determines whether a mode has already arrived, whereas $\mathcal I_\alpha^-$ determines whether an arrived mode can still affect the subsequent QPC evolution. 

\begin{figure*}[t]
    \centering
    \includegraphics[width=0.8\textwidth]{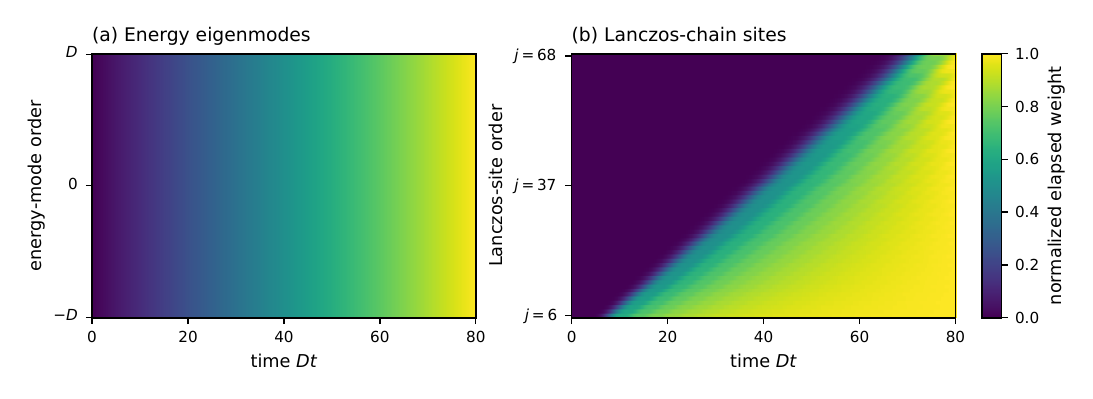}
    \caption{
        Elapsed coupling weights $\mathcal I_\alpha^+$ for (a) energy eigenmodes and (b) chain sites. For visibility, the weight of each mode is normalized by its value at $T$. Energy eigenmodes with nonzero contact overlap accumulate weight from the beginning of the evolution, whereas chain sites are reached successively by the propagating contact disturbance.
    }
    \label{fig:elapsed_weight_basis_comparison}
\end{figure*}

\begin{figure*}[t]
    \centering
    \includegraphics[width=0.8\textwidth]{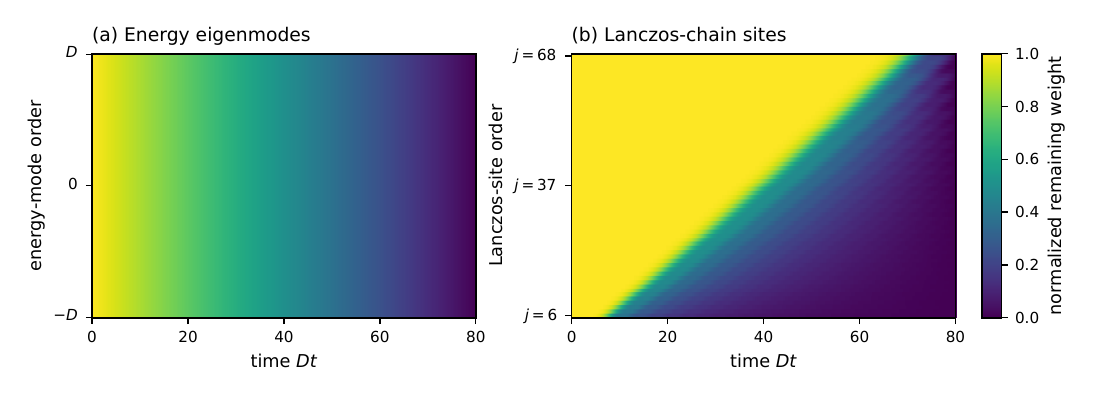}
    \caption{
        Remaining coupling weights $\mathcal I_\alpha^-$ for (a) energy eigenmodes and (b) chain sites. For visibility, the weight of each mode is normalized by its value at $t=0$. For energy eigenmodes the remaining weight decreases throughout the simulation interval. Chain sites exhibit an ordered departure, but dispersive tails broaden the interval over which their remaining
        coupling is appreciable.
    }
    \label{fig:remaining_weight_basis_comparison}
\end{figure*}

The usefulness of these criteria depends crucially on the reservoir basis. To keep the number of simultaneously retained modes small, different reservoir modes should couple appreciably to the QPC over different time intervals. The energy eigenbasis illustrates why this does not happen automatically. For an energy eigenmode $\hat\kappa_{\alpha,\epsilon}^\dagger$ of a time-independent lead, free evolution changes the contact amplitude only by a phase, so that $\left|\chi_\alpha[\kappa_{\alpha,\epsilon};t]\right|^2 = \left| \chi_\alpha[\kappa_{\alpha,\epsilon};0] \right|^2$.
Consequently:
\begin{align}
\mathcal I_\alpha^+ [\kappa_{\alpha,\epsilon};t] 
&= t\, \left| \chi_\alpha[\kappa_{\alpha,\epsilon};0] \right|^2 
\\
\nonumber 
\mathcal I_\alpha^- [\kappa_{\alpha,\epsilon};t,T] 
&= (T-t)\, \left|\chi_\alpha[\kappa_{\alpha,\epsilon};0]\right|^2.
\end{align}
Thus all energy modes have the same time dependence of their elapsed and remaining coupling weights, differing only in their overall coupling strengths. Their coupling to the QPC extends throughout the entire simulation interval rather than being concentrated over different portions of it. The energy basis therefore does not provide the successive arrival and departure of reservoir modes needed to maintain a small moving window, as illustrated in Figs.~\ref{fig:elapsed_weight_basis_comparison}(a) and \ref{fig:remaining_weight_basis_comparison}(a).

A chain representation introduces a natural arrival order. Because the QPC couples only to the first site, increasingly distant chain sites become coupled to the QPC only as the propagation front reaches them. Their elapsed coupling weights therefore remain small until that time, producing the sequence of arrivals shown in Fig.~\ref{fig:elapsed_weight_basis_comparison}(b). The number of chain sites involved in the dynamics therefore grows progressively with time. This sequential propagation underlies the usefulness of chain mappings in open-system and tensor-network calculations \cite{Chin2010,Tamascelli2018}.

However, the departure structure is less sharp. After the propagating front has passed a chain site, its remaining coupling weight decreases, but dispersive tails can keep $\mathcal I_\alpha^-$ appreciable over a broad time interval, as shown in Fig.~\ref{fig:remaining_weight_basis_comparison}(b). Thus, the chain basis provides a clear sequence of arrivals, whereas the departures are spread over broader time intervals, causing the number of retained sites to grow with time.

This suggests choosing a reservoir basis in which the coupling of different modes to the QPC is concentrated over different parts of the simulation interval, so that only a finite moving window of modes needs to be retained at any given time.

\begin{figure*}[t]
    \centering
    \includegraphics[width=0.98\textwidth]{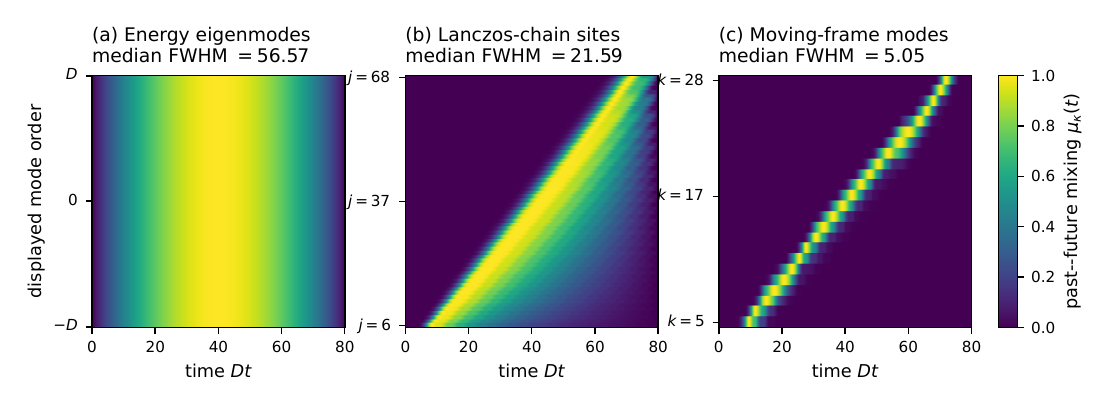}
    \caption{
        Comparison of normalized $\frac{\mathcal I_\alpha^+[\kappa;t]\mathcal I_\alpha^-[\kappa;t,T]}{\left[\mathcal I_\alpha^+[\kappa;T]\right]^2}$ (see Appendix~\ref{sec:temporal_mode_localization} Eq.~ \eqref{eq:past_future_mixing}) in three reservoir representations for a rectangular contact spectrum. Energy eigenmodes remain mixed over almost the full interval. Lanczos chain sites acquire an ordered incoming front but retain broad future tails. The corrected tape-recorder moving-frame modes form narrow, sequential active windows. The bulk median FWHM is $56.57$, $21.59$, and $5.05$, respectively. Parameters are $N=256$, $T=80$, $D=1$, $\Delta t=0.001$, $r_{\mathrm{cut}}=10^{-6}$, and $D\tau_{\min}=1$.}
    \label{fig:temporal_localization_bases}
\end{figure*}

Figure~\ref{fig:temporal_localization_bases} compares energy eigenmodes, chain sites, and moving-frame modes for the same normalized rectangular contact spectrum. A quantitative comparison of their temporal localization, including the tails of representative modes on a logarithmic scale, is given in Appendix~\ref{sec:temporal_mode_localization}. 

\subsection{Contact wave packets and coupling-weight operators}

We now construct such a basis from the elapsed and remaining coupling weights. To this end, we represent the weights in Eqs.~\eqref{eq:forward_weight}, \eqref{eq:backward_weight} by positive-semidefinite operators in the single-particle Hilbert space of each lead. This allows the reservoir modes to be selected directly from the spectra of the corresponding coupling-weight operators.

Because lead $\alpha$ is noninteracting, the free evolution of the orbital at the QPC-lead contact is a single-particle problem. In the interaction picture with respect to $\hat H_\alpha$:
\begin{equation}
\hat a_{\alpha,0}^{\dagger}(t) = e^{i\hat H_\alpha t} \hat a_{\alpha,0}^{\dagger} e^{-i\hat H_\alpha t} = \sum_{j\geq0} \alpha_{\alpha,j}(t) \hat a_{\alpha,j}^{\dagger}.
\label{eq:contact_evolution}
\end{equation}
The coefficients $\alpha_{\alpha,j}(t)$ obey the single-particle Schrödinger equation generated by $\hat H_\alpha$, with $\alpha_{\alpha,j}(0)=\delta_{j0}$. They are the amplitudes of the normalized contact wave packet:
\begin{align}
\ket{\alpha_\alpha(t)} & = \hat a_{\alpha,0}^{\dagger}(t)\ket{\tilde 0}_\alpha = \sum_{j\geq 0} \alpha_{\alpha,j}(t)\ket{j}_\alpha, 
\\ \nonumber 
\ket{j}_\alpha & = \hat a_{\alpha,j}^{\dagger}\ket{\tilde 0}_\alpha.
\label{eq:contact_wavepacket}
\end{align}

For a normalized reservoir mode, written in chain basis:
\begin{equation}
\hat\kappa_\alpha^\dagger = \sum_{j\geq0} \kappa_{\alpha,j}\hat a_{\alpha,j}^\dagger, \quad \ket{\kappa_\alpha} = \hat\kappa_\alpha^\dagger\ket{\tilde 0}_\alpha, \quad \braket{\kappa_\alpha}{\kappa_\alpha}=1,
\end{equation}
the instantaneous coupling amplitude~\eqref{eq:mode_coupling_amplitude} is simply the overlap:
\begin{equation}
\chi_\alpha[\kappa;t] = \braket{\alpha_\alpha(t)}{\kappa_\alpha}.
\label{eq:mode_coupling_overlap}
\end{equation}
Thus, the coupling weight of a reservoir mode over a given time interval is determined by its overlap with the contact wave packet over that interval. The corresponding coupling weights over the elapsed and remaining intervals are represented by the positive-semidefinite operators:
\begin{align}
\hat\rho_{+,\alpha}(t)
&= \int_0^t d\tau\, \ket{\alpha_\alpha(\tau)} \bra{\alpha_\alpha(\tau)},
\label{eq:forward_lc}
\\
\hat\rho_{-,\alpha}(t;T)
&= \int_t^T d\tau\, \ket{\alpha_\alpha(\tau)} \bra{\alpha_\alpha(\tau)}.
\label{eq:backward_lc}
\end{align}
Their expectation values reproduce the scalar coupling weights introduced in Eqs.~\eqref{eq:forward_weight} and \eqref{eq:backward_weight}:
\begin{equation}
\begin{gathered}
\mathcal I_\alpha^+[\kappa;t] = \bra{\kappa_\alpha} \hat\rho_{+,\alpha}(t) \ket{\kappa_\alpha} 
\\ 
\mathcal I_\alpha^-[\kappa;t,T] = \bra{\kappa_\alpha} \hat\rho_{-,\alpha}(t;T) \ket{\kappa_\alpha}.
\label{eq:coupling_weight_expectation}
\end{gathered}
\end{equation}
Finally, since the elapsed and remaining intervals partition the interval $[0,T]$, the two operators satisfy:
\begin{equation}
\hat\rho_{-,\alpha}(t;T) = \hat\rho_{+,\alpha}(T) - \hat\rho_{+,\alpha}(t).
\label{eq:coupling_weight_operator_partition}
\end{equation}

The two operators therefore provide the ingredients for constructing the two boundaries of the moving reservoir window. The spectrum of $\hat\rho_{+,\alpha}(t)$ determines when reservoir modes must enter the propagated state, while $\hat\rho_{-,\alpha}(t;T)$ will determine when modes that have already entered can be removed. We consider these two constructions in turn.

\subsection{Incoming modes from accumulated coupling weight}
We first construct the incoming boundary of the moving window using the minimal-forward-light-cone procedure of Refs.~\cite{ArefyevaPolyakov2026,Polyakov2022Bandlimited}, adapted here to the coupling-weight operator \eqref{eq:forward_lc}. The eigenvectors of $\hat\rho_{+,\alpha}(t)$ form an orthogonal set of reservoir modes, with the corresponding eigenvalues giving their accumulated coupling weights:
\begin{equation}
\begin{gathered}
\hat\rho_{+,\alpha}(t) \ket{\kappa_{\alpha,p}^{+}(t)} = \mathcal I_{\alpha,p}^{+}(t) \ket{\kappa_{\alpha,p}^{+}(t)}, \\ \nonumber \mathcal I_{\alpha,1}^{+}(t) \geq   \mathcal I_{\alpha,2}^{+}(t)\geq\cdots.
\label{eq:forward_eigenproblem}
\end{gathered}
\end{equation}
We regard an eigenmode as appreciably coupled over the elapsed interval when:
\begin{equation}
\mathcal I_{\alpha,p}^{+}(t) > r_{\mathrm{cut}}\, \mathcal I_{\alpha,1}^{+}(t).
\label{eq:forward_significance}
\end{equation}
Thus, the eigenmodes satisfying \eqref{eq:forward_significance} span the reservoir subspace with appreciable accumulated coupling up to time $t$. For the bounded spectral densities considered here, the largest forward coupling weight remains bounded as the simulation interval is extended. Appendix~\ref{sec:saturation} gives the corresponding uniform bound and proves convergence to a finite limit.

The eigenvectors of $\hat\rho_{+,\alpha}(t)$ vary with time and are therefore not used directly as the fermionic modes of the many-body propagation. Instead, they are used to construct a fixed set of incoming modes and their arrival times. For a prescribed simulation interval $[0,T]$, the construction proceeds by a recursive backward sweep. 

At $t=T$, we retain the eigenmodes satisfying Eq.~\eqref{eq:forward_significance}. As the sweep proceeds backward, $\hat\rho_{+,\alpha}(t)$ is diagonalized within the subspace of modes whose arrival times have not yet been assigned. Denoting the largest and smallest eigenvalues in this subspace by $\mathcal I_{\alpha,\max}^{+}(t)$ and $\mathcal I_{\alpha,\min}^{+}(t)$, respectively, we test the condition:
\begin{equation}
\mathcal I_{\alpha,\min}^{+}(t) < r_{\mathrm{cut}}\, \mathcal I_{\alpha,\max}^{+}(t).
\label{eq:incoming_freeze_condition}
\end{equation}
When this condition is first satisfied during the backward sweep, the corresponding eigenvector is fixed as an incoming mode $\ket{\kappa_{\alpha,p}^{\mathrm{in}}}$, and that time is recorded as its arrival time $t_{\alpha,p}^{\mathrm{in}}$. This mode is excluded from subsequent rotations, and the sweep continues in the remaining subspace. Modes that remain above the cutoff at $t=0$ are assigned zero arrival time.

\begin{figure*}[t]
    \centering
    \includegraphics[width=0.98\textwidth]{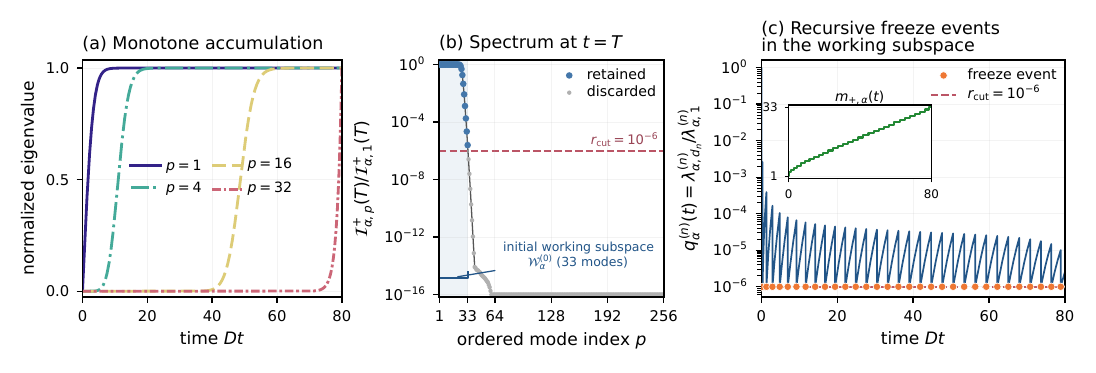}
    \caption{
        Construction of the fixed incoming basis.
        (a) Four ordered eigenvalues of the accumulated coupling-weight operator $\hat\rho_{+,\alpha}(t)$, each normalized by its value at $T$. Different ordered eigenvalues acquire their accumulated weight at different times, providing the temporal structure to construct the incoming modes. 
        (b) Spectrum of $\hat\rho_{+,\alpha}(T)$ normalized by its largest eigenvalue;
        the blue points correspond to modes above $r_{\mathrm{cut}}$ and are retained at the start of the backward sweep; the gray points lie below the cutoff.
        (c) Recursive backward construction. The blue curve shows the ratio of the smallest to the largest accumulated coupling weight within the subspace whose arrival times have not yet been assigned. Whenever this ratio falls below $r_{\mathrm{cut}}$ (orange markers), the corresponding eigenmode is fixed as an incoming mode and its arrival time is recorded. The sweep then continues in the remaining subspace. The vertical jump shows the ratio after the selected mode is removed, and the next threshold test is performed at the next earlier time sample.  The inset shows the number $m_{+,\alpha}(t)$ of incoming modes present at forward time $t$. Parameters are $N=256$, $T=80$, $\Delta t=0.001$, $D=1$, and $r_{\mathrm{cut}}=10^{-6}$.
        The normalizations in panels (a) and (b) are used only for display; the construction use the corresponding unnormalized operators.
    }
    \label{fig:incoming_mode_backward_sweep}
\end{figure*}

In forward propagation, mode $\hat\kappa_{\alpha,p}^{\mathrm{in}}$ enters the propagated state at $t=t_{\alpha,p}^{\mathrm{in}}$. Thus the backward construction delays the arrival of the reservoir modes as much as allowed by the prescribed coupling-weight criterion. The number of incoming modes that have arrived by time $t$ is:
\begin{equation}
m_{+,\alpha}(t) = \sum_p \Theta\!\left(t-t_{\alpha,p}^{\mathrm{in}}\right).
\end{equation}
Figure~\ref{fig:incoming_mode_backward_sweep} illustrates this construction.

\subsection{Outgoing modes from remaining coupling weight}

The incoming construction determines when reservoir modes enter the propagated state. We now determine when modes that have already arrived can be removed. The relevant quantity is the remaining coupling weight over the interval $[t,T]$: a mode can be discarded once its remaining coupling to the QPC becomes negligible on the prescribed accuracy scale.

At time $t$, an outgoing mode must be chosen from the span of the incoming modes that have already arrived and have not yet been removed. We refer to this span as the active subspace of lead $\alpha$ and denote its orthogonal projector by $\hat P_{\alpha}^{\mathrm{act}}(t)$. The remaining coupling-weight operator restricted to this subspace is:
\begin{equation}
\hat\rho_{-,\alpha}^{\mathrm{act}}(t;T) = \hat P_{\alpha}^{\mathrm{act}}(t) \hat\rho_{-,\alpha}(t;T) \hat P_{\alpha}^{\mathrm{act}}(t).
\label{eq:projected_backward_lc}
\end{equation}
Its eigenvectors are orthogonal combinations of the currently active modes, ordered by their remaining coupling weights:
\begin{align}
\hat\rho_{-,\alpha}^{\mathrm{act}}(t;T) \ket{\bar\kappa_{\alpha,p}(t)} &= \mathcal I_{\alpha,p}^{-}(t;T) \ket{\bar\kappa_{\alpha,p}(t)},
\\ \nonumber
\mathcal I_{\alpha,1}^{-}(t;T) &\geq\cdots\geq \mathcal I_{\alpha,r_\alpha(t)}^{-}(t;T),
\label{eq:backward_eigenproblem}
\end{align}
where $r_\alpha(t)$ is the number of active modes. The eigenmode with the smallest eigenvalue therefore has the least remaining coupling to the QPC and is the natural candidate for removal.

During forward propagation, we monitor the smallest remaining coupling weight relative to the largest one \cite{ArefyevaPolyakov2026}. An outgoing event occurs when:
\begin{equation}
\mathcal I_{\alpha,\min}^{-}(t;T) < r_{\mathrm{cut}}\, \mathcal I_{\alpha,\max}^{-}(t;T).
\label{eq:escape_condition}
\end{equation}
The first time at which this condition is satisfied defines the departure time $t_{\alpha,p}^{\mathrm{out}}$. The corresponding eigenvector is fixed as the outgoing mode $\ket{\kappa_{\alpha,p}^{\mathrm{out}}}$ and removed from the active subspace. Subsequent outgoing modes are selected from the remaining active subspace, so that previously removed outgoing modes cannot reenter the propagated state.

Together, the incoming and outgoing constructions define the moving active subspace. Incoming modes enlarge this subspace as their accumulated coupling becomes appreciable, while outgoing modes remove combinations whose remaining coupling has become negligible. If $m_{-,\alpha}(t)$ denotes the number of outgoing modes removed by time $t$, the number of active reservoir modes is:
\begin{equation}
r_\alpha(t) = m_{+,\alpha}(t)-m_{-,\alpha}(t).
\label{eq:relevant_number}
\end{equation}
Thus reservoir modes can continue to arrive while combinations that no longer affect the subsequent QPC evolution are removed, allowing the explicitly propagated reservoir modes to form a finite moving window whose size can remain bounded as propagation continues.

\subsection{Wave-function evolution in the active basis}
The incoming and outgoing constructions determine the active reservoir subspace at each time. We now describe the many-body evolution within this moving subspace. The fixed incoming modes with $t_{\alpha,p}^{\mathrm{in}}\leq t$ form an orthonormal basis for the reservoir modes that have arrived. Within their span, the outgoing construction separates the modes removed up to time $t$ from their orthogonal complement, which defines the current active subspace. Choosing an orthonormal basis adapted to this decomposition defines a time-dependent unitary change of one-particle basis, denoted by $W_\alpha(t)$. This change of representation is exact: it preserves the canonical anticommutation relations, while its time dependence generates the frame term introduced below.

Let $\{\hat\kappa_{\alpha,p}^{\mathrm{act}}\}_{p=1}^{r_\alpha(t)}$ be an orthonormal basis of the active subspace. The component of the contact annihilation operator supported in this subspace is:
\begin{equation}
\hat a_{\alpha,0}^{\mathrm{act}}(t)
\equiv \sum_{p=1}^{r_\alpha(t)}
\chi_{\alpha,p}^{\mathrm{act}}(t)
\hat\kappa_{\alpha,p}^{\mathrm{act}},\quad\chi_{\alpha,p}^{\mathrm{act}}(t) = \left\{ \hat a_{\alpha,0}(t), \hat\kappa_{\alpha,p}^{\mathrm{act}\dagger} \right\}.
\label{eq:contact_projection}
\end{equation}
Equation~\eqref{eq:contact_projection} is the orthogonal projection of the contact field onto the active subspace. The reservoir approximation consists of replacing $\hat a_{\alpha,0}(t)$ by $\hat a_{\alpha,0}^{\mathrm{act}}(t)$ in the QPC-lead interaction. The omitted component $ \hat a_{\alpha,0}(t) - \hat a_{\alpha,0}^{\mathrm{act}}(t)$ contains modes that have not yet arrived and modes that have already departed, whose couplings are controlled by the elapsed and remaining coupling-weight criteria, respectively. The moving-basis transformation itself is exact.

Between discrete arrival and departure events, $r_\alpha(t)$ is constant, although the orthonormal active vectors may rotate smoothly. The motion of the active basis is characterized by the Hermitian connection:
\begin{equation}
\begin{gathered}
\xi_{\alpha,pq}(t) = i\bra{\kappa_{\alpha,p}^{\mathrm{act}}(t)} \partial_t \ket{\kappa_{\alpha,q}^{\mathrm{act}}(t)} = \\ = \left[iW_\alpha^\dagger(t)\dot W_\alpha(t)\right]_{pq}. 
\label{eq:frame_generator}
\end{gathered}
\end{equation}
Expressing the wavefunction in coordinates that co-move with this basis adds $-iW_\alpha^\dagger\dot W_\alpha=-\xi_\alpha$ to the one-particle generator. After second quantization, the same basis motion gives the quadratic term
$-\sum_{pq}\xi_{\alpha,pq}
\hat\kappa_{\alpha,p}^{\mathrm{act}\dagger}
\hat\kappa_{\alpha,q}^{\mathrm{act}}$.

In the interaction picture with respect to the free leads, the free reservoir dynamics is contained in the time-dependent contact amplitudes $\chi_{\alpha,p}^{\mathrm{act}}(t)$. The trajectory therefore evolves between arrival and departure events under:
\begin{equation}
\begin{gathered}
\hat H_{\mathrm{eff}}(t) =
\hat{\tilde H}_S(t) \,\, + \\ + \sum_{\alpha=L,R} \sum_{p=1}^{r_\alpha(t)} \left[\hat V_\alpha^\dagger\chi_{\alpha,p}^{\mathrm{act}}(t) \hat\kappa_{\alpha,p}^{\mathrm{act}} + \chi_{\alpha,p}^{\mathrm{act}*}(t)
\hat\kappa_{\alpha,p}^{\mathrm{act}\dagger} \hat V_\alpha \right] - 
\\
- \sum_{\alpha=L,R} \sum_{p,q=1}^{r_\alpha(t)} \xi_{\alpha,pq}(t) \hat\kappa_{\alpha,p}^{\mathrm{act}\dagger} \hat\kappa_{\alpha,q}^{\mathrm{act}}.
\label{eq:effective_hamiltonian}    
\end{gathered}
\end{equation}
Here $\hat{\tilde H}_S(t)$ is the QPC Hamiltonian, while the second term is the QPC-lead interaction after projection of the contact field onto the active subspace. The final quadratic term is kinematic: it accounts for the time dependence of the basis used to represent the active reservoir modes. It introduces no additional physical reservoir interaction, and is retained exactly. The only approximation in Eq.~\eqref{eq:effective_hamiltonian} is the omission of the orthogonal contact component $\hat a_{\alpha,0}(t)-\hat a_{\alpha,0}^{\mathrm{act}}(t)$.

At a departure event, the moving basis isolates the outgoing combination as a single fermionic mode. Once its residual QPC coupling has been omitted, this mode is absent from the subsequent truncated Hamiltonian. It may, however, remain entangled with the QPC and the active reservoir modes through correlations established before departure. Removing it from the propagated Fock space therefore requires a partial trace.

For the outgoing mode $\hat\kappa_{\alpha,p}^{\mathrm{out}}$, the two occupation sectors are selected by
\begin{equation}
\hat P_{\alpha,p}^{(0)} = 1 -    \hat\kappa_{\alpha,p}^{\mathrm{out}\dagger}\hat\kappa_{\alpha,p}^{\mathrm{out}}, \quad\hat P_{\alpha,p}^{(1)} = \hat\kappa_{\alpha,p}^{\mathrm{out}\dagger}
\hat\kappa_{\alpha,p}^{\mathrm{out}}.
\end{equation}
Using the occupation basis of this mode, a trajectory state immediately before removal can be written as:
\begin{equation}
|\Psi\rangle = |\phi_0\rangle\otimes|0\rangle_{\alpha,p} + |\phi_1\rangle\otimes|1\rangle_{\alpha,p}, \quad |\phi_n\rangle = {}_{\alpha,p}\!\langle n|\Psi\rangle ,
\end{equation}
where $|\phi_n\rangle$ is an unnormalized state of the QPC and the remaining active modes. Its norm gives the weight of occupation sector $n$,
\begin{equation}
p_n = \langle\phi_n|\phi_n\rangle = \langle\Psi| \hat P_{\alpha,p}^{(n)} |\Psi\rangle .
\end{equation}
Tracing out the outgoing mode gives:
\begin{align}
\hat\rho_{\mathrm{rem}} = &\operatorname{Tr}_{\alpha,p} |\Psi\rangle\langle\Psi| = \sum_{n=0}^{1} |\phi_n\rangle\langle\phi_n| = \sum_{n=0}^{1} p_n|\psi_n\rangle\langle\psi_n|, 
\\ \nonumber
& |\psi_n\rangle = \frac{|\phi_n\rangle}{\sqrt{p_n}},
\quad p_n>0.
\label{eq:outgoing_partial_trace_unraveling}
\end{align}
Rather than propagate the mixed state $\hat\rho_{\mathrm{rem}}$, each Monte Carlo trajectory samples one occupation sector with probability $p_n$, removes the outgoing mode from its Fock-space representation, and continues with the normalized conditional state $|\psi_n\rangle$. For any observable $\hat O$ acting only on the QPC and the remaining active modes:
\begin{equation}
\operatorname{Tr} \left(\hat O\hat\rho_{\mathrm{rem}}\right) = \sum_{n=0}^{1} p_n \langle\psi_n|\hat O|\psi_n\rangle .
\label{eq:outgoing_ensemble_average}
\end{equation}
Thus occupation sampling is an exact stochastic unraveling of the partial trace \cite{Polyakov2022RealTime}. It introduces neither a physical measurement of the QPC nor an additional approximation. The approximation has already been made when the residual QPC coupling of the outgoing mode was omitted; the stochastic sampling only provides a pure-state representation of the resulting reduced dynamics.

\subsection{Full-state error of a single outgoing-mode truncation}

A departure event introduces one approximation: the residual QPC coupling to the selected outgoing mode is omitted. We quantify the resulting full-state error before the mode is traced out. Consider an outgoing mode $\hat\kappa_{\alpha,p}^{\mathrm{out}}$ selected at the departure time $t_d=t_{\alpha,p}^{\mathrm{out}}$. For $s\in[t_d,T]$, the contact field has the exact decomposition:
\begin{align}
\hat a_{\alpha,0}(s) &= \hat a_{\alpha,0}^{\perp}(s) + \eta_{\alpha,p}(s) \hat\kappa_{\alpha,p}^{\mathrm{out}}, \\ \nonumber \eta_{\alpha,p}(s) &= \left\{ \hat a_{\alpha,0}(s),\hat\kappa_{\alpha,p}^{\mathrm{out}\dagger}\right\},
\label{eq:discarded_contact_component}
\end{align}
where $\{\hat a_{\alpha,0}^{\perp}(s), \hat\kappa_{\alpha,p}^{\mathrm{out}\dagger}\}=0$. The contribution of this mode to the QPC-lead interaction is:
\begin{equation}
\Delta\hat H_{\alpha,p}(s) = \hat V_\alpha^\dagger \eta_{\alpha,p}(s) \hat\kappa_{\alpha,p}^{\mathrm{out}}
+ \eta_{\alpha,p}^{*}(s) \hat\kappa_{\alpha,p}^{\mathrm{out}\dagger} \hat V_\alpha .
\label{eq:residual_interaction}
\end{equation}
We denote by $\hat H(s)$ the Hamiltonian retaining this residual coupling. Omitting it gives the truncated Hamiltonian:
\begin{equation}
\hat H_\perp(s) = \hat H(s)-\Delta\hat H_{\alpha,p}(s).
\label{eq:single_event_truncated_hamiltonian}
\end{equation}
Both Hamiltonians act on the same full QPC-reservoir Hilbert space. In the truncated evolution, the outgoing mode therefore remains present as a decoupled degree of freedom; no partial trace has yet been taken.

Let $|\Psi(s)\rangle$ and $|\Psi_\perp(s)\rangle$ be the full states generated by $\hat H(s)$ and $\hat H_\perp(s)$, respectively, from the common state at the departure time,
\begin{equation}
|\Psi(t_d)\rangle = |\Psi_\perp(t_d)\rangle \equiv |\Psi_d\rangle . \label{eq:single_event_common_state}
\end{equation}
We quantify the error introduced by omitting the residual coupling through the full-state infidelity:
\begin{equation}
I_{\alpha,p}(s) = 1 - \left| \langle\Psi_\perp(s)|\Psi(s)\rangle\right|.
\label{eq:truncation_infidelity}
\end{equation}
The partial trace over the outgoing mode is performed only after this truncation and therefore does not enter the definition of $I_{\alpha,p}(s)$.

The scalar coefficient $\eta_{\alpha,p}(\tau)$ in Eq.~\eqref{eq:residual_interaction} is instantaneous coupling amplitude of the outgoing mode to the QPC. The coupling weight omitted between the departure time $t_d$ and a later time $s$ is therefore:
\begin{equation}
\varepsilon_{\alpha,p}(s) = \int_{t_d}^{s} d\tau\, |\eta_{\alpha,p}(\tau)|^2, \quad t_d\leq s\leq T.
\label{eq:postdeparture_weight}
\end{equation}
At $s=T$, $\varepsilon_{\alpha,p}(T) =\mathcal I_\alpha^-[\kappa_{\alpha,p}^{\mathrm{out}};t_d,T]$, so it is precisely the remaining coupling weight used in the departure criterion. This scalar weight is not itself a state error: the response also depends on the QPC operators and on propagation under the truncated Hamiltonian.

Let $\hat U_\perp(s,\tau)$ denote the propagator generated by $\hat H_\perp$. For a square-integrable state-vector-valued input $F(\tau)$, we define the finite-interval response factors as the smallest constants for which:
\begin{equation}
\begin{aligned}
\left\|\int_{t_d}^{s} d\tau\,\hat U_\perp(s,\tau) \hat V_\alpha^\dagger F(\tau) \right\| 
&\leq C_{\alpha,-}(s) \left(  \int_{t_d}^{s} d\tau\, \|F(\tau)\|^2 \right)^{1/2} \\ 
\left\| \int_{t_d}^{s} d\tau\, \hat U_\perp(s,\tau) \hat V_\alpha F(\tau) \right\|
&\leq C_{\alpha,+}(s) \left( \int_{t_d}^{s} d\tau\, \|F(\tau)\|^2 \right)^{1/2}
\end{aligned}
\label{eq:finite_response_factors}
\end{equation}
These are the induced $L^2$-to-Hilbert-space norms of the two retarded response maps. They convert the norm of the discarded interaction envelope into a state-vector response at time $s$.

The derivation is given in Appendix~\ref{sec:error_bounds}. Let $E_{\alpha,-}(s)$ and $E_{\alpha,+}(s)$ be the squared $L^2$ norms of the annihilation and creation sources in the exact Duhamel identity. The fermionic relations $\hat\kappa^\dagger\hat\kappa=\hat N_\kappa$ and $\hat\kappa\hat\kappa^\dagger=1-\hat N_\kappa$ give the exact sum $E_{\alpha,-}(s)+E_{\alpha,+}(s) = \varepsilon_{\alpha,p}(s)$.

For $|\delta\Psi_{\alpha,p}(s)\rangle =|\Psi(s)\rangle-|\Psi_\perp(s)\rangle$, the response bounds and the Cauchy-Schwarz inequality between the two source channels yield:
\begin{equation}
\|\delta\Psi_{\alpha,p}(s)\|\leq \sqrt{C_{\alpha,-}^2(s)+C_{\alpha,+}^2(s)}\, \sqrt{\varepsilon_{\alpha,p}(s)}.
\label{eq:single_event_state_response_bound}
\end{equation}
No assumption on the occupation of the outgoing mode is required. 

For normalized states, $1-|\langle\phi|\psi\rangle| \leq\dfrac{1}{2}\|\,|\phi\rangle-|\psi\rangle\,\|^2$. Therefore the infidelity is bounded:
\begin{equation}
I_{\alpha,p}(s) \leq \frac{1}{2} \left[ C_{\alpha,-}^2(s)+C_{\alpha,+}^2(s)\right] \varepsilon_{\alpha,p}(s).
\label{eq:infidelity_bound}
\end{equation}
This bound follows from the exact propagator identity and does not use an expansion in the residual interaction.

Unitarity of $\hat U_\perp(s,\tau)$ and the Cauchy-Schwarz inequality in time give $C_{\alpha,\pm}(s)\leq \sqrt{s-t_d}\,\|\hat V_\alpha\|$. Equation~\eqref{eq:infidelity_bound} then implies:
\begin{equation}
I_{\alpha,p}(s) \leq (s-t_d) \|\hat V_\alpha\|^2\varepsilon_{\alpha,p}(s).
\label{eq:infidelity_bound_linear}
\end{equation}
At $s=T$, the selected outgoing mode belongs to the lowest eigenspace of the active-subspace remaining-weight operator at $t_d$. Hence:
\begin{equation}
\begin{gathered}
\varepsilon_{\alpha,p}(T) = \mathcal I_\alpha^-[\kappa_{\alpha,p}^{\mathrm{out}};t_d,T] = 
\\
= \,\mathcal I_{\alpha,r_\alpha(t_d)}^-(t_d;T) < r_{\mathrm{cut}}\, \mathcal I_{\alpha,1}^-(t_d;T).
\label{eq:single_event_departure_weight}
\end{gathered}
\end{equation}
Substitution gives:
\begin{align}
I_{\alpha,p}(T)
&< \frac{r_{\mathrm{cut}}}{2} \left[ C_{\alpha,-}^2(T)+C_{\alpha,+}^2(T) \right] \mathcal I_{\alpha,1}^-(t_d;T),
\label{eq:single_event_threshold_response_bound}
\\
I_{\alpha,p}(T)
&< (T-t_d) \|\hat V_\alpha\|^2 r_{\mathrm{cut}}\, \mathcal I_{\alpha,1}^-(t_d;T).
\label{eq:single_event_threshold_explicit_bound}
\end{align}
The bound applies separately to each outgoing-mode removal, with the departure time, remaining coupling weight, and response factors evaluated for that event.

\subsection{Computational scaling of the active window}
\label{sec:active_window_scaling}

The calculations use three independent numerical controls, which act at different levels of the representation. The relative threshold $r_{\mathrm{cut}}$ controls the one-particle reservoir construction: it determines incoming arrivals and, in the threshold construction, outgoing departures. The per-lead cap $r_{\alpha}^{\mathrm{cap}}$ fixes the maximum number of active ring modes used by the production solver. The cutoff $N_{\mathrm{qp}}^{\max}$ restricts the total quasiparticle-number sectors of the joint QPC-reservoir Fock space. These parameters have distinct roles and are varied independently.

The threshold construction above defines departure events through Eq.~\eqref{eq:escape_condition}. The transport calculations retain the same forward and incoming construction but use a fixed-rank realization of the outgoing boundary. When an arrival would increase the active rank above $r_{\alpha}^{\mathrm{cap}}$, the remaining coupling-weight operator is diagonalized in the active subspace and the least-future-coupled mode is selected as outgoing. Thus the cap determines when a production departure is triggered, whereas the remaining coupling-weight operator determines which active mode is removed.

The cap is increased until currents, occupations, and other ensemble-averaged observables are converged. For each cap-selected outgoing mode, the single-event estimate in Eq.~\eqref{eq:infidelity_bound} applies with the actual remaining coupling weight of that mode. The numerical saturation of the threshold-selected active width, discussed below, explains why a time-independent converged cap can remain sufficient over long propagation intervals.

For given values of the rank caps, the solver represents the two QPC quasiparticle modes and the preallocated ring modes of both leads in one joint fermionic Fock space. With
\begin{equation}
M_{\mathrm{alloc}} = 2+r_L^{\mathrm{cap}}+r_R^{\mathrm{cap}},
\end{equation}
and a cutoff $N_{\mathrm{qp}}^{\max}$ on the total number of quasiparticles in this joint space, the propagated state-vector dimension is:
\begin{equation}
\begin{gathered}
D_{\mathrm{traj}} \left( r_L^{\mathrm{cap}},r_R^{\mathrm{cap}},N_{\mathrm{qp}}^{\max}\right)
= \\ = 
\sum_{n=0}^{\min\{N_{\mathrm{qp}}^{\max},M_{\mathrm{alloc}}\}} \binom{M_{\mathrm{alloc}}}{n} \leq 2^{M_{\mathrm{alloc}}}.
\label{eq:trajectory_dimension}
\end{gathered}
\end{equation}
For the symmetric implementation, $r_L^{\mathrm{cap}}=r_R^{\mathrm{cap}}$. At fixed $N_{\mathrm{qp}}^{\max}$, the leading growth is $M_{\mathrm{alloc}}^{N_{\mathrm{qp}}^{\max}}/ N_{\mathrm{qp}}^{\max}!$, rather than exponential in the allocated number of modes.

The total-occupation cutoff is effective because the active region is open to quasiparticle flux. The anomalous terms in $\hat{\tilde H}_S(t)$ create and annihilate quasiparticle pairs at the QPC. Created quasiparticles propagate into the leads and leave the propagated state when outgoing modes are sampled and removed. In a stationary or periodically stationary regime, injection and escape can establish a stationary active-region occupation distribution. We therefore choose $N_{\mathrm{qp}}^{\max}$ by increasing it until currents, occupations, and other ensemble-averaged observables are converged.

Reservoir-basis preprocessing, finite-chain and time-step errors, and the number of stochastic trajectories provide additional costs or uncertainties that are separate from the three controls above and from $D_{\mathrm{traj}}$.

\begin{figure}[h!]
    \centering
    \includegraphics[width=0.4\textwidth]{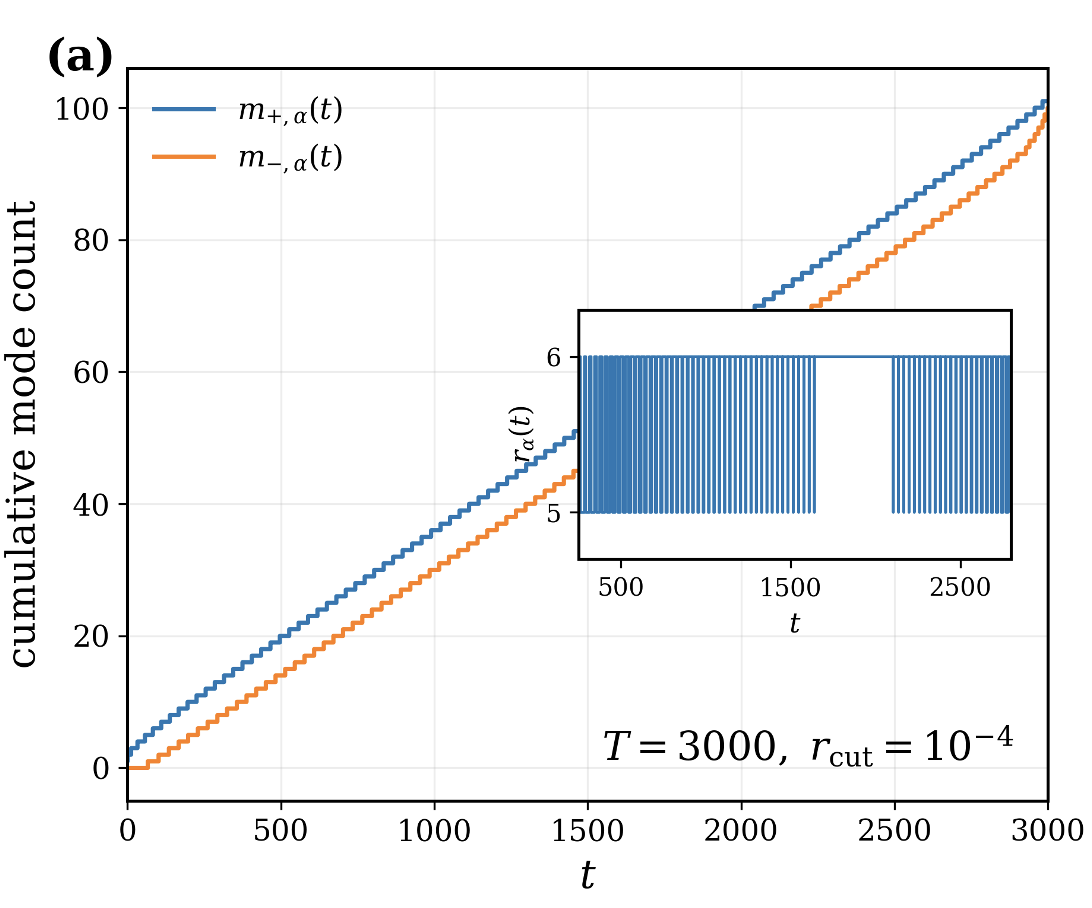}
    \hfill
    \includegraphics[width=0.4\textwidth]{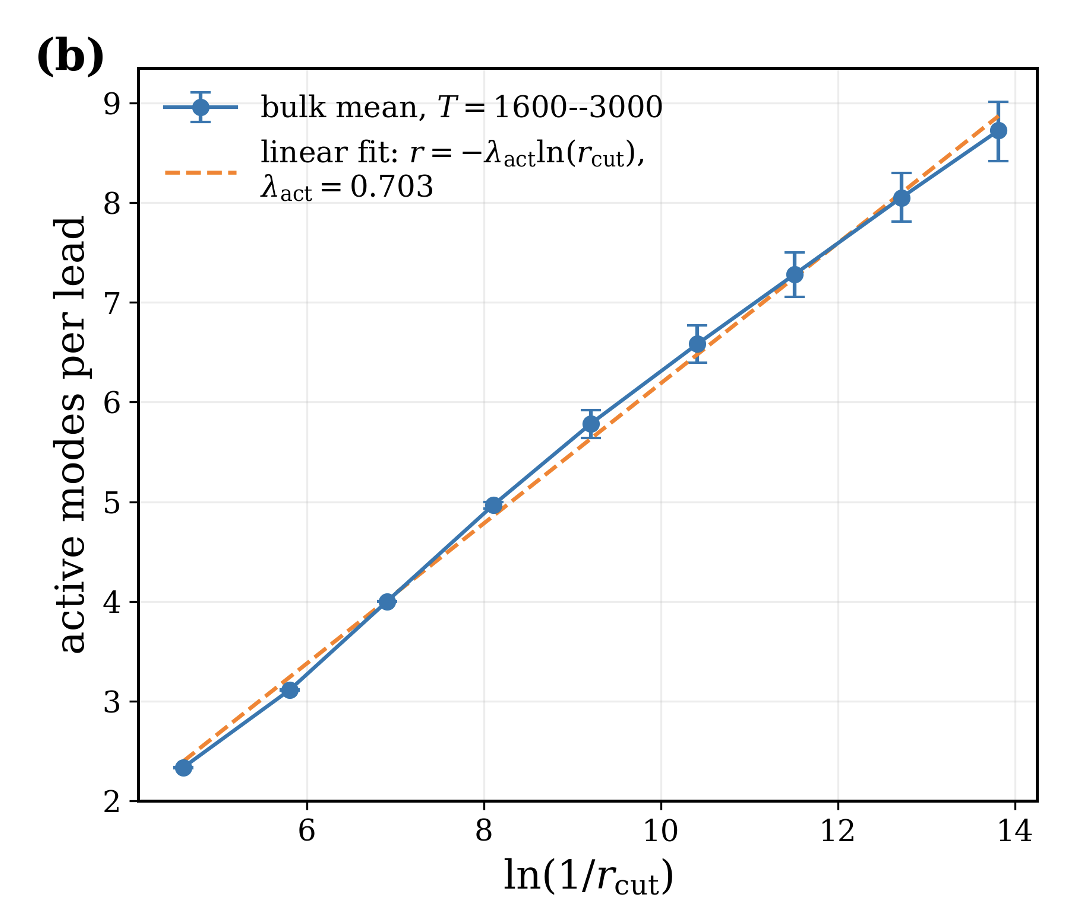}
    \caption{
        Threshold-defined active-window scaling for one lead. (a) Cumulative incoming and outgoing mode counts at $T=3000$ and $r_{\mathrm{cut}}=10^{-4}$. Their difference $r_\alpha(t)=m_{+,\alpha}(t)-m_{-,\alpha}(t)$ remains between five and six through the bulk of the interval, as shown in the inset. (b) Bulk mean active width, averaged over $T=1600$, $2200$, and $3000$, versus $\ln(1/r_{\mathrm{cut}})$. Error bars span the values obtained for these three values of $T$, and the dashed line is the fit specified in the legend. The same relative threshold is used in the incoming and outgoing constructions. The chain length is chosen so that the ballistic time to the boundary is approximately $2T$. Increasing the chain length leaves the active count unchanged in the validation case; halving the time step leaves its bulk mode and maximum unchanged.
    }
    \label{fig:active_window_threshold_scaling}
\end{figure}

The threshold-defined geometry can be calibrated independently of the production rank caps. Figure~\ref{fig:active_window_threshold_scaling} shows that the cumulative incoming and outgoing mode counts both grow throughout the simulated interval, whereas their difference remains confined to a narrow band through its bulk. For $T=3000$ and $r_{\mathrm{cut}}=10^{-4}$, the active width in one lead alternates between five and six over most of the interval. Averaging the bulk width over $T=1600$, $2200$, and $3000$ gives an approximately logarithmic threshold dependence with fitted logarithmic coefficient $\lambda_{\mathrm{act}}=0.703$. This is an empirical finite-interval calibration of the threshold construction; the fixed-rank production realization remains subject to an independent convergence test.

\subsection{Contact spectral density of the active reservoir}
\label{sec:active_window_contact_spectrum}

The continuum-to-chain mapping and the subsequent active-window reduction are distinct approximations to the reservoir representation. To separate them, we compare the target contact spectral density $J_{\mathrm{target}}(\omega)$, the spectrum $J_{\mathrm{chain}}(\omega)$ reconstructed from the full finite Lanczos chain, and the spectrum $J_{\mathrm{aw}}(\omega)$ reconstructed from the moving active window. At each time, the contact vector and its freely propagated image are transformed to the moving reservoir basis and restricted to the current active interval. Their overlap defines the active-window return amplitude $C_{\mathrm{aw}}(t)$. For $x\in\{\mathrm{chain},\mathrm{aw}\}$, we use the same undamped finite-time transform:
\begin{equation}
J_x^{(T)}(\omega) = 2g^2\operatorname{Re} \int_0^T dt\, e^{i\omega t}C_x(t),
\label{eq:active_window_contact_spectrum}
\end{equation}
where $g^2=(2\pi)^{-1}\int d\omega\,J_{\mathrm{target}}(\omega)$ fixes the contact normalization. No artificial broadening or spectral window is applied. Spectral deviations are quantified by the normalized $L^1$ distance:
\begin{equation}
E(A,B) = \frac{\int_{-2}^{2}d\omega\,|A(\omega)-B(\omega)|}{\int_{-2}^{2}d\omega\,|B(\omega)|}.
\label{eq:contact_spectral_error}
\end{equation}

\begin{figure*}[t]
    \centering
    \includegraphics[width=0.8\textwidth]{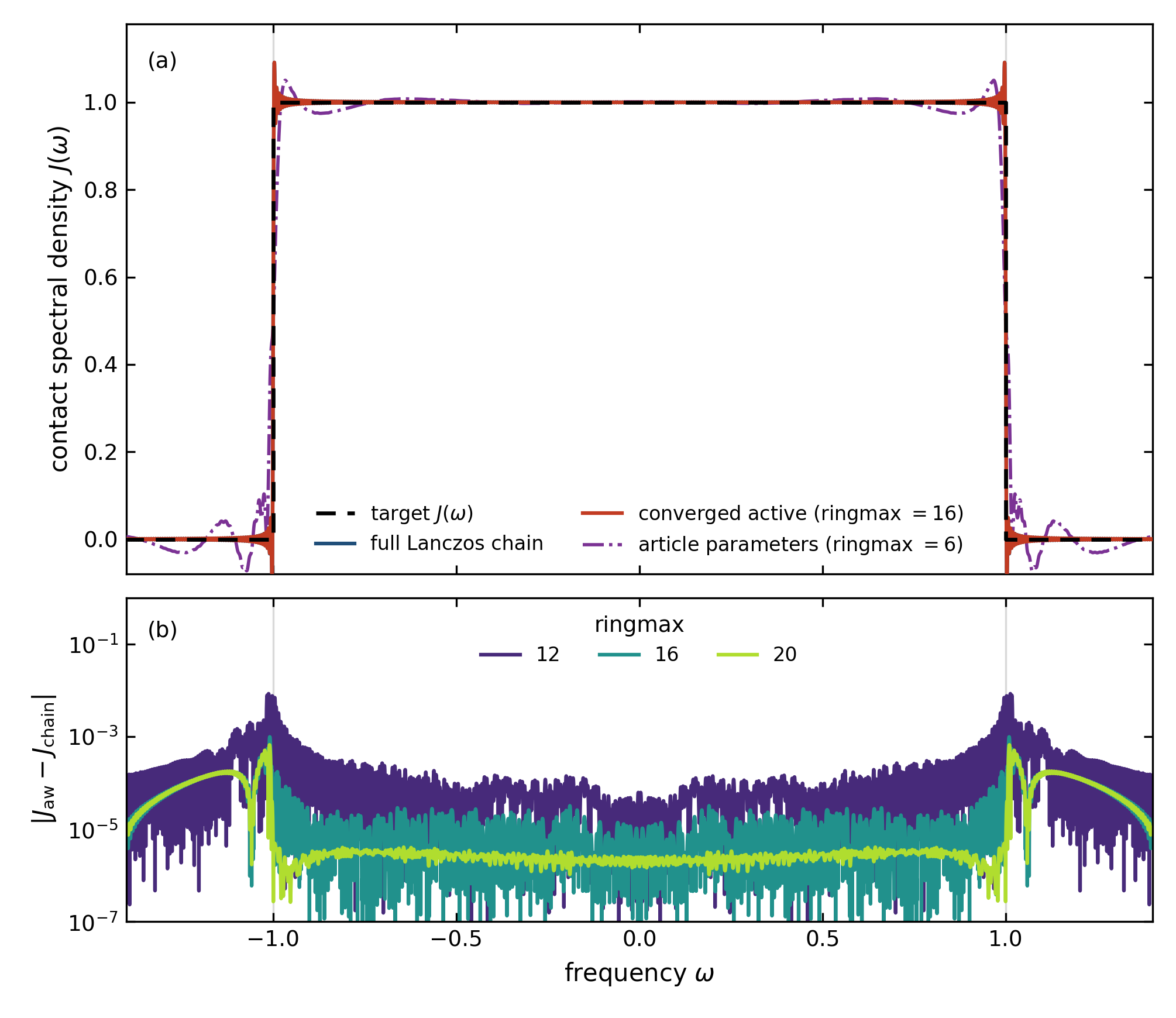}
    \caption{
        Contact spectral density represented by the moving active-window reservoir for a unit-height flat band with edges at $|\omega|=1$. (a) Ideal target density (black dashed), the undamped finite-time transform from the full $2000$-site Lanczos chain (blue), the converged active-window reconstruction at $T=1000$, $\Delta t=0.05$, $r_{\mathrm{cut}}=10^{-6}$, and $\mathtt{ringmax}=16$ (red), and the reconstruction at the parameters used in the flat-band transport calculation, $T=500$, $\Delta t=0.1$, $r_{\mathrm{cut}}=10^{-5}$, and $\mathtt{ringmax}=6$ (purple dash-dotted). (b) Absolute deviation of the active-window spectrum from the matching full-chain transform at $T=1000$ for $\mathtt{ringmax}=12$, $16$, and $20$. No artificial broadening or spectral window is applied. The normalized $L^1$ active-window error is $7.9\times10^{-5}$ at $\mathtt{ringmax}=16$, compared with $2.70\times10^{-2}$ for the article-parameter curve relative to its own matching full-chain transform.
    }
    \label{fig:active_window_contact_spectrum}
\end{figure*}

At the converged settings in Fig.~\ref{fig:active_window_contact_spectrum}, the active-window spectrum differs from the matching full-chain transform by $7.9\times10^{-5}$ in normalized $L^1$. Increasing $\mathtt{ringmax}$ from $16$ to $20$ changes this value only to $6.9\times10^{-5}$. The remaining deviation from the ideal rectangular target is approximately $3.3\times10^{-3}$ and is dominated by truncating the undamped Fourier integral at finite $T$. The edge oscillations in Fig.~\ref{fig:active_window_contact_spectrum}(a) therefore do not indicate additional broadening by the active window. At the parameters used in the flat-band transport calculation, the active-window spectrum differs from its matching full-chain transform by $2.70\times10^{-2}$. This quantity calibrates the reservoir representation and is not, by itself, an estimate of the error in a transport observable.

Having characterized the active-window geometry and calibrated the contact spectral representation, we now turn to transport benchmarks.

\section{Results}\label{sec:results}

\subsection{Lorentzian-reservoir benchmark}
\label{sec:lorentzian_validation}

We first benchmark the tape-recorder method for reservoirs with a Lorentzian spectral density. The corresponding reservoir correlation function can be represented accurately by a moderate number of exponential terms, making this case comparatively favorable for HEOM \cite{Jin2008}. It therefore provides a direct comparison with an independent numerical approach under conditions in which both calculations can be converged.

In the tape-recorder calculation, each continuum reservoir is first represented by a discretized spectral measure and then mapped unitarily to a nearest-neighbor chain by Lanczos tridiagonalization. The contact spectral accuracy of both the finite-chain and active-window representations is quantified in Sec.~\ref{sec:active_window_contact_spectrum}.

\begin{figure*}[!htbp]
    \centering
    \includegraphics[width=0.8\linewidth]{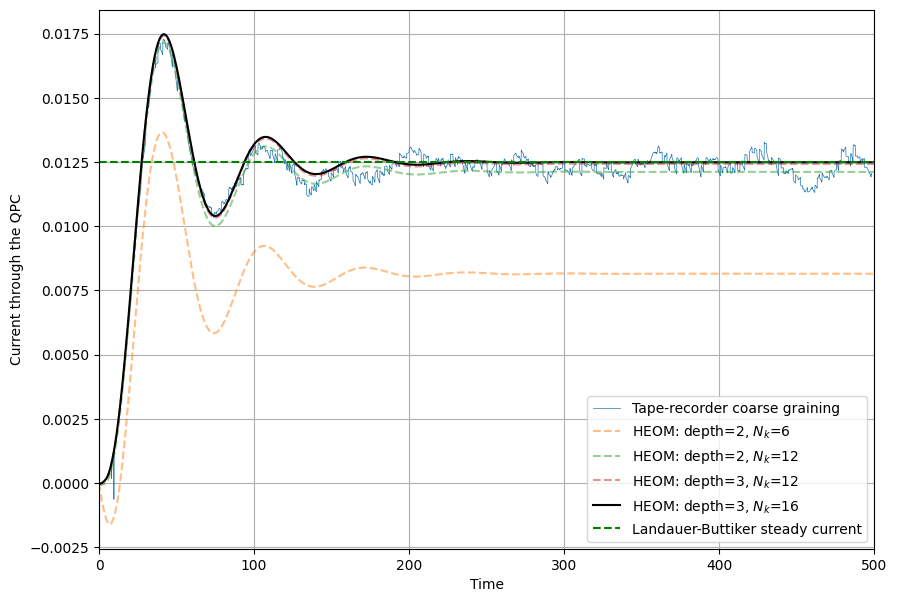}
    \caption{
        Current through the QPC for Lorentzian reservoirs. The tape-recorder coarse-graining result is compared with HEOM calculations for several values of the hierarchy depth and of the number \(N_k\) of exponential terms in the bath-correlation decomposition. The horizontal dashed line denotes the Landauer-Büttiker steady-state current. The \(N_k=6\) calculation is not converged, whereas the \(N_k=12\) and \(N_k=16\) results agree closely with the tape-recorder calculation and with the steady-state reference.
    }
    \label{fig:lorentzian_heom_comparison}
\end{figure*}

Figure~\ref{fig:lorentzian_heom_comparison} compares the transient current obtained with the tape-recorder method and HEOM. The Landauer-Büttiker steady-state current provides the long-time reference. For \(N_k=6\), the HEOM current approaches a stationary value that differs from this reference, indicating that the exponential representation of the reservoir correlation function is not converged. Increasing \(N_k\) to \(12\) substantially improves the result. The \(N_k=12\) and \(N_k=16\) curves closely follow the tape-recorder dynamics and approach the Landauer-Büttiker value at long times.

At fixed \(N_k=12\), increasing the hierarchy depth from \(2\) to \(3\) produces only a negligible change. This indicates that, for the displayed parameters, the HEOM result is already converged with respect to the hierarchy depth and that the remaining error is controlled primarily by the exponential decomposition of the bath correlation function. Once this decomposition is sufficiently accurate, the tape-recorder, HEOM, and Landauer-Büttiker results are mutually consistent. This comparison therefore serves as a sanity check for both the tape-recorder evolution and the Lanczos chain construction, although it does not by itself establish accuracy for more difficult spectral densities.

\subsection{Flat-band reservoirs}
\label{sec:flat_band_results}

We next consider the more demanding case of reservoirs with the flat spectral density
\begin{equation}
    J(\omega)
    =
    \begin{cases}
        J_0, & |\omega| < \omega_0,\\
        0,   & |\omega| \geq \omega_0.
    \end{cases}
    \label{eq:flat_spectral_density}
\end{equation}
The discontinuities at $\omega=\pm\omega_0$ produce slowly decaying temporal correlations and make the corresponding correlation function considerably more difficult to approximate by a finite sum of exponentials. The contact spectral errors of the finite Lanczos chain and the active-window representation are quantified independently in Sec.~\ref{sec:active_window_contact_spectrum}; here we focus on the resulting transport dynamics.

\begin{figure*}
    \centering
    \subfloat[
        Long-time current for flat-band reservoirs. Increasing the hierarchy depth from $2$ to $3$ at fixed $N_k=18$ has little effect, whereas changing $N_k$ substantially modifies the HEOM result. None of the displayed HEOM calculations is converged to the Landauer-Büttiker steady-state current.
        \label{fig:flat_heom_comparison}
    ]{\includegraphics[width=0.8\textwidth]{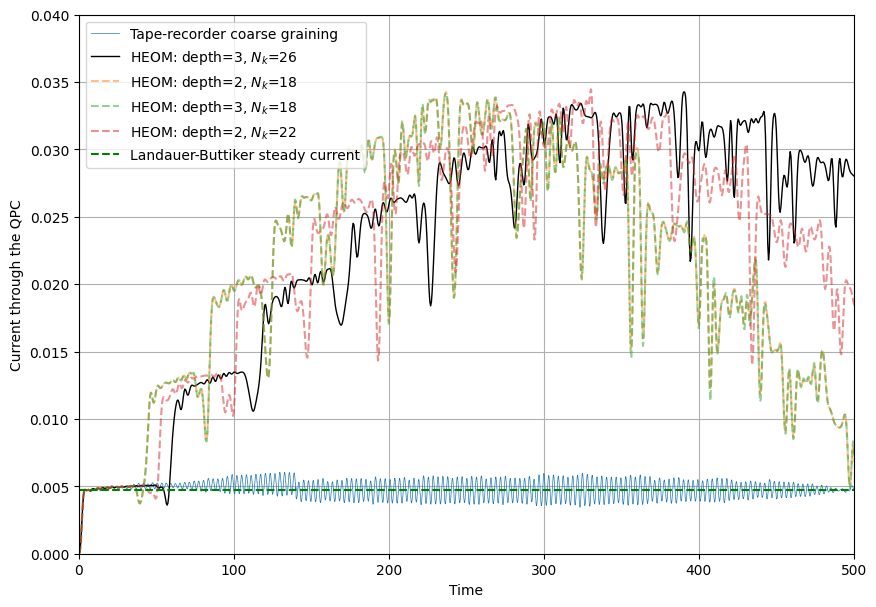}
    }
\vspace{0.8em}

    \subfloat[
        Comparison with direct Schrödinger-equation integration using explicitly truncated leads. The vertical dashed line marks the estimated onset of the finite-size revival.
        \label{fig:flat_lead_truncation}
    ]{\includegraphics[width=0.8\textwidth]{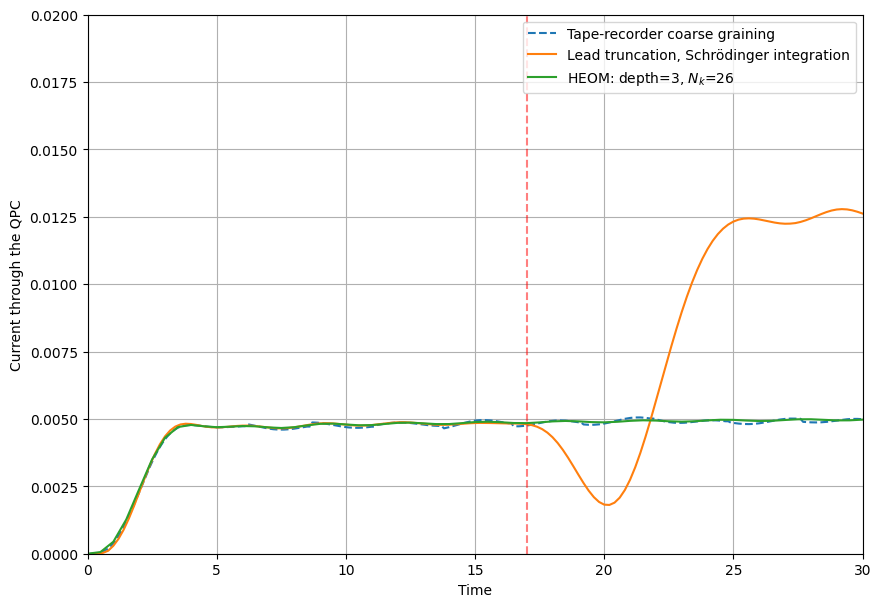}
    }
    \caption{
        Transport validation for flat-band reservoirs. Panel (a) compares the long-time current with HEOM and the Landauer-Büttiker steady-state result. Panel (b) compares the short-time dynamics with direct Schrödinger evolution for explicitly truncated leads. The contact spectral representation of the same reservoir is calibrated in Sec.~\ref{sec:active_window_contact_spectrum}.
    }
    \label{fig:flat_band_validation}
\end{figure*}

The long-time current is shown in
Fig.~\ref{fig:flat_band_validation}(a). The tape-recorder calculation rapidly approaches a quasi-stationary current close to the Landauer-Büttiker value and remains near it throughout the simulated interval. In contrast, the HEOM trajectories depend strongly on the number \(N_k\) of retained exponential terms. At fixed \(N_k=18\), the results for hierarchy depths \(2\) and \(3\) are nearly indistinguishable, while increasing \(N_k\), changes the HEOM trajectory substantially. This behavior shows that the exponential representation of the reservoir correlation function, rather than the hierarchy-depth truncation, is the dominant source of error in the displayed HEOM calculations. Even the result with \(N_k=26\) develops large irregular deviations and does not converge to the Landauer-Büttiker current over the considered time interval. With high enough $N_k$ HEOM will surely converge, however our point here is that the computational resources required for HEOM converge in this case are much higher that required for tape-recorder convergence.

A short-time validation is also provided in Fig.~\ref{fig:flat_band_validation}(b), where the tape-recorder result is compared with direct Schrödinger equation integration for explicitly truncated leads. The finite leads used in this calculation consist of the first few sites of the same Lanczos chains employed in the tape-recorder simulation. The two calculations agree closely before the finite-size revival time, which verifies that the tape-recorder construction reproduces the local dynamics of the explicitly represented part of the chain.

The short-time comparison verifies the local dynamics generated by the mapped chain before finite-size recurrences, while the Landauer-Büttiker comparison tests the long-time stationary transport. The HEOM curves probe a separate numerical issue: for a sharply bounded spectral density, convergence of the exponential reservoir representation is substantially more demanding than in the Lorentzian case.

\subsection{Interacting and driven QPC models}
\label{sec:interacting_driven_qpc}

\begin{figure*}[t]
    \centering
    \subfloat[
    Current through the QPC as a function of the on-site potential \(\varepsilon_d\) for an interacting QPC. Two current peaks appear, separated approximately by \(\Delta \varepsilon_d = U\), which is the characteristic Coulomb-blockade splitting.
    \label{fig:coulomb_blockade_qpc}
    ]{\includegraphics[width=0.47\textwidth]{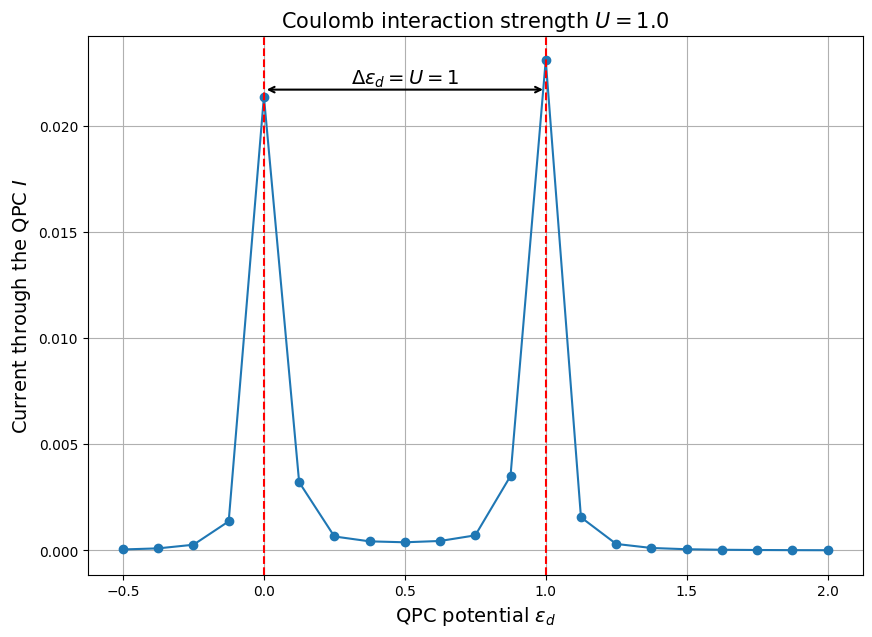}}
    \hfill
    \subfloat[Current as a function of \(A/\omega\) for a periodically driven QPC with \(\varepsilon_d(t)=A\cos(\omega t)\), shown for several Coulomb interaction strengths \(U\). The vertical dashed line marks the first zero of the Bessel function \(J_0\), where the current is strongly suppressed.
    \label{fig:driven_qpc_bessel}
    ]{\includegraphics[width=0.47\textwidth]{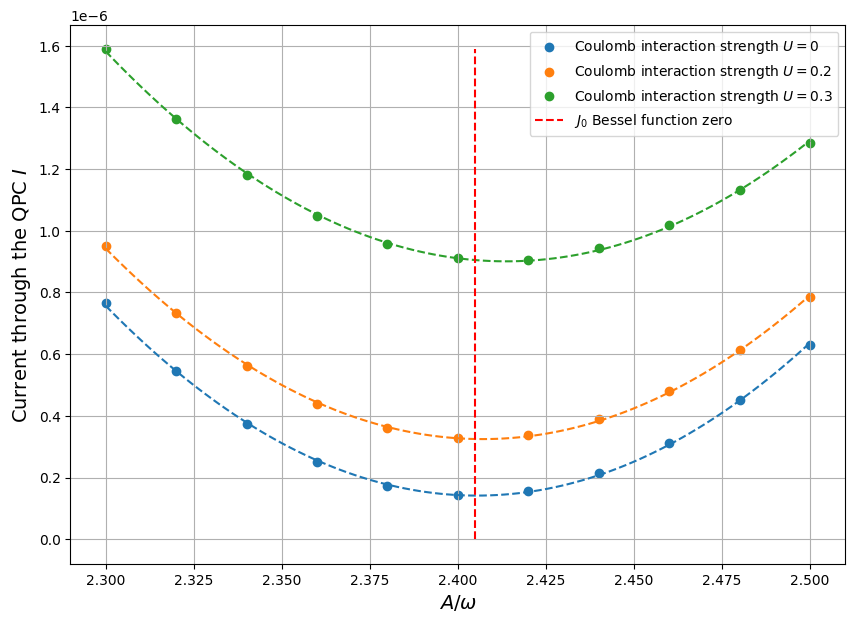}}
    \caption{
        Examples of interacting and driven QPC models accessible to the tape-recorder coarse-graining method.
    }
    \label{fig:interacting_driven_qpc}
\end{figure*}
Beyond the benchmark cases considered above, the tape-recorder coarse-graining method also allows us to study more demanding QPC models that include interaction effects and explicit time dependence. In this subsection we illustrate this point with two examples: Coulomb blockade in an interacting QPC and current suppression in a periodically driven QPC.

We first consider a QPC with Coulomb interaction and sweep the on-site potential \(\varepsilon_d\) through the transport window. The result is shown in Fig.~\ref{fig:interacting_driven_qpc}(a). In the presence of a charging interaction \(U\), the current exhibits two distinct resonant peaks rather than a single resonance. The separation of the peaks is approximately equal to the interaction strength,
\begin{equation}
    \Delta \varepsilon_d \approx U,
\end{equation}
which is the characteristic signature of Coulomb blockade \cite{Kouwenhoven1997}. Physically, the first peak corresponds to transport through the empty QPC level, while the second peak corresponds to transport through the interaction-shifted level. This demonstrates that the method correctly captures interaction-induced spectral splitting and can be used to simulate transport regimes beyond the noninteracting limit.

As a second example, we consider a periodically driven QPC with
\begin{equation}
\varepsilon_d(t) = A \cos(\omega t),
\end{equation}
and study the current as a function of the dimensionless driving amplitude \(A/\omega\). The results are presented in Fig.~\ref{fig:interacting_driven_qpc}(b) for several values of the Coulomb interaction strength \(U\). In the noninteracting case, it is known analytically that periodic driving renormalizes the effective tunneling by a Bessel factor \(J_0(A/\omega)\). As a consequence, the current is strongly suppressed when \(A/\omega\) approaches a zero of the Bessel function, corresponding to coherent destruction of tunneling \cite{Grossmann1991}. The first zero,
\begin{equation}
x_0 \approx 2.4048,
\end{equation}
is indicated by the vertical dashed line in Fig.~\ref{fig:interacting_driven_qpc}(b). The simulated current indeed develops a pronounced minimum in the vicinity of this point.

The same calculation can be repeated in the interacting case, where the Coulomb term modifies the transport but a simple analytical treatment is not available. Figure~\ref{fig:interacting_driven_qpc}(b) shows that the current-suppression effect persists for finite \(U\), while the overall magnitude and curvature of \(I(A/\omega)\) become interaction dependent. This show that tape-recorder coarse graining not only reproduces a known analytically tractable effect for \(U=0\), but also extends straightforwardly to the interacting regime where analytical results are not readily available.

\subsection{Comparison with analytical benchmarks}
\label{sec:analytical_qpc}
In addition to the numerical benchmarks discussed above, the tape-recorder coarse-graining method can be validated against independent theoretical approaches available in limiting cases of the present model. These comprise an exact Floquet Green-function calculation for the noninteracting driven system and a diagrammatic nonequilibrium Green-function calculation within the self-consistent second-Born approximation. 

\begin{figure*}[t]
    \centering
    \subfloat[
    Current as a function of \(A/\Omega\) for a periodically driven QPC with \(\varepsilon_d(t)=A\cos(\Omega t)\). Comparison with the exact Floquet Green-function solution. The vertical dashed line marks the first zero of the Bessel function \(J_0\).
    \label{fig:analytical_dr_qpc}
    ]{\includegraphics[width=0.47\linewidth]{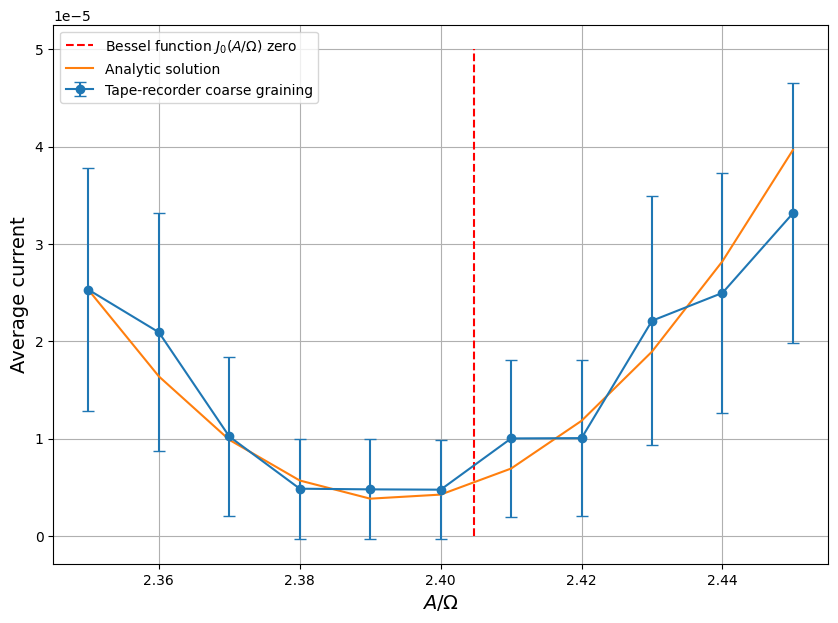}}
    \hfill
    \subfloat[Current as a function of the interaction strength $U$ for a static ($A=0$) QPC. Comparison with the Keldysh nonequilibrium Green-function calculation within the self-consistent second-Born approximation.
    \label{fig:analytical_int_qpc}
    ]
    {
    \includegraphics[width=0.47\linewidth]{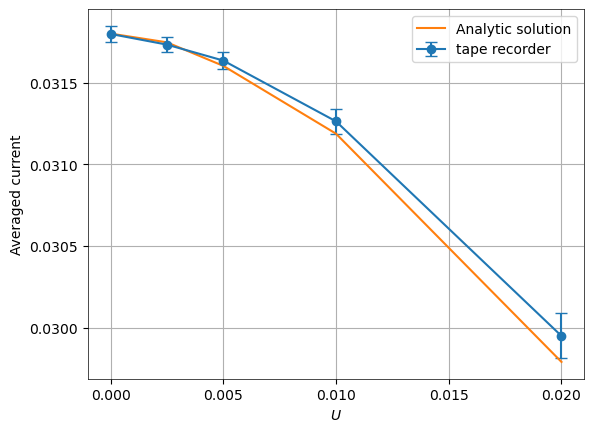}}
    \caption{
        Comparison of the tape-recorder coarse-graining method with independent Green-function benchmarks. In both panels, the tape-recorder results are averaged over 100 independent Monte Carlo trajectories, and the error bars indicate the corresponding statistical uncertainty.
    }
    \label{fig:analytical_qpc}
\end{figure*}

We first consider the periodically driven noninteracting QPC ($U=0$). In this limit, the transport problem admits an exact Floquet Green-function solution \cite{Kohler2005}. Figure~\ref{fig:analytical_qpc}(a) compares the tape-recorder results with the exact Floquet solution for the current as a function of the dimensionless driving amplitude $A/\Omega$. The tape-recorder data were obtained by averaging over 100 independent Monte Carlo trajectories. The two calculations are indistinguishable within the statistical uncertainty of the Monte Carlo sampling throughout 
the displayed parameter range. In particular, the tape-recorder method reproduces the suppression of the current near the first zero of the Bessel function $J_0(A/\Omega)$, confirming that it correctly describes coherent photon-assisted transport in the driven noninteracting limit.

A second benchmark is provided by the static interacting model. In the absence of periodic driving, the transport current can be computed using the self-consistent second-Born approximation \cite{HaugJauho1996,Myohanen2009}, which is expected to be reliable in the weak-interaction regime. As shown in Fig.~\ref{fig:analytical_qpc}(b), the tape-recorder current agrees closely with the second-Born result over the displayed interaction range. No systematic deviation is observed within the numerical accuracy of the calculations.

Taken together, these comparisons validate the tape-recorder coarse-graining method against two independent Green-function benchmarks: the exact Floquet Green-function solution for periodically driven transport and the Keldysh nonequilibrium Green-function calculation within the self-consistent second-Born approximation. They complement the numerical benchmarks presented above and provide additional evidence for the accuracy of the proposed method.

\section{Discussion}\label{sec:discussion}

The numerical results test several distinct components of the tape-recorder construction. For Lorentzian reservoirs, the agreement with converged HEOM calculations verifies the transient propagation in a case for which the bath correlation function is efficiently represented by exponentials. Agreement with the Landauer-Büttiker current provides an independent test of the long-time noninteracting limit. This benchmark is deliberately favorable to HEOM and shows that the tape-recorder calculation reproduces an established result when both approaches are converged.

The flat-band calculation probes a different regime. The discontinuity of the spectral density produces an algebraically decaying correlation function and makes a finite exponential representation substantially more demanding. The agreement with direct Schrödinger propagation before the finite-chain revival verifies the short-time dynamics generated by the Lanczos-mapped reservoir. The agreement of the long-time current with the Landauer-Büttiker value tests the combined effect of the chain representation and the active-window reduction. In the displayed HEOM calculations, changing the number $N_k$ of exponential terms has a much larger effect than increasing the hierarchy depth. The dominant convergence control for these parameters is therefore the exponential representation of the reservoir correlation function. The separate contact-spectrum calculation further distinguishes the finite-chain representation from the active-window reduction and shows that the latter can be converged independently by increasing the active rank.

The principal difference from a spatial lead truncation is the criterion used to retain environmental degrees of freedom. A finite chain keeps every site up to an artificial boundary, regardless of whether the corresponding combinations of states can still influence the central region. Its maximum reliable propagation time therefore increases with chain length. Tape-recorder coarse graining instead reorganizes the lead into modes that are localized in time. A mode enters the active window when its accumulated coupling becomes significant and leaves when its remaining coupling over the rest of the prescribed simulation interval falls below the relative threshold. At fixed $r_{\mathrm{cut}}$, the cumulative numbers of incoming and outgoing modes continue to grow, whereas their difference remains in a narrow band over the simulated intervals. Tightening the threshold increases this active width approximately as $\ln(1/r_{\mathrm{cut}})$.

The relationship to tensor-network and influence-functional methods is complementary. Methods such as TEMPO compress a temporal influence functional, with their efficiency governed by its singular-value spectrum or temporal entanglement \cite{Strathearn2018}. Tape-recorder coarse graining first changes the physical reservoir basis so that incoming, active, and outgoing physical modes are separated explicitly. The time-dependent rotation and the associated basis-rotation couplings are retained in the propagated Hamiltonian. The approximation consists in omitting the contact-field component outside the active subspace. The resulting finite-mode problem may itself be propagated using a dense solver, a tensor network, or a stochastic trajectory method. Combining the adaptive reservoir basis with a more scalable many-body propagator is therefore a natural route to larger interacting devices.

The outgoing-mode occupations also provide a non-Markovian trajectory interpretation. For a structured reservoir, a mode remains in the active subspace while it can still mediate memory back to the QPC. Its occupation is sampled only after its remaining coupling over the simulation interval falls below the threshold. This avoids imposing an instantaneous-jump approximation on a mode that still participates in the non-Markovian dynamics. Tracing over the outgoing mode and stochastic sampling of its occupation are two representations of the same truncated reduced dynamics.

Several limitations remain. The present construction uses single-particle propagation in quadratic reservoirs. Interacting leads would require a many-body generalization of the coupling-weight construction. The calculations considered here also use a zero-temperature, maximal-bias filled/empty initial condition. General Gaussian reservoir states require the corresponding particle and hole purification channels, whose numerical cost should be assessed separately. The finite Lanczos chain, time step, active-rank cap, total-quasiparticle cutoff, and number of stochastic trajectories provide independent numerical controls. Finally, the analytical full-state bound is event resolved: it is evaluated separately for each outgoing-mode removal using the departure time and remaining coupling weight of that event.

\section{Conclusion}\label{sec:conclusion}

We have introduced a fermionic tape-recorder coarse-graining method for the real-time simulation of interacting devices coupled to structured, noninteracting reservoirs. The method reorganizes each lead into orthogonal incoming, active, and outgoing modes. A mode enters the active window when its accumulated coupling to the device becomes significant, remains active while it can affect the subsequent dynamics, and leaves when its remaining coupling over the prescribed simulation interval falls below the relative threshold. The basis rotation, including the induced basis-rotation couplings, is retained exactly. The approximation is the omission of the residual QPC coupling outside the active subspace.

For each outgoing-mode truncation, we derive a nonperturbative finite-interval bound on the infidelity between the exact and truncated full QPC-reservoir states. The bound is expressed through the remaining coupling weight of the outgoing mode and finite-interval response factors of the retained dynamics. The fermionic bound does not require an assumption about the occupation of the outgoing mode. For an essentially bounded contact spectral density, the largest eigenvalue of the forward coupling-weight operator remains bounded as the simulation interval is extended. Numerically, the number of simultaneously active modes saturates over the examined intervals at fixed $r_{\mathrm{cut}}$ and grows approximately logarithmically as the threshold is reduced. Long-time propagation therefore does not require retaining a number of reservoir modes proportional to the elapsed time.

For a two-site quantum point contact, the method agrees with converged HEOM and Landauer-Büttiker results for Lorentzian reservoirs. For flat-band reservoirs, it agrees with direct Schrödinger propagation before finite-size recurrences and reproduces the noninteracting stationary current, while the tested HEOM results remain sensitive to the exponential reservoir representation. Independent Green-function comparisons validate the noninteracting driven limit and provide an additional benchmark in the static interacting regime. The same framework captures Coulomb-blockade peak splitting and drive-induced current suppression in interacting and periodically driven QPC models.

These results show that a moving window of physical reservoir modes provides a practical wave-function representation for long-time fermionic transport. Natural extensions include general Gaussian reservoir states, multichannel geometries, larger interacting devices, and the use of tensor-network propagators within the active reservoir basis. 

The corresponding numerical implementation is available in \cite{QPCCode}.

\subsection*{Acknowledgements}
The work of N. A. was supported by the Theoretical Physics and Mathematics Advancement Foundation “BASIS” Grant No. 23-2-2-26-1.

\appendix
\section{Temporal localization of reservoir modes}\label{sec:temporal_mode_localization}
The elapsed and remaining weights divide the complete coupling history,
\begin{equation}
\mathcal I_\alpha^{\mathrm{tot}}[\kappa;T] = \mathcal I_\alpha^+[\kappa;t] + \mathcal I_\alpha^-[\kappa;t,T] = \mathcal I_\alpha^+[\kappa;T].
\label{eq:total_weight_partition}
\end{equation}
To compare modes with different total coupling strengths, we normalize the two parts,
\begin{equation}
\begin{gathered}
p_{\alpha,\kappa}^{+}(t) = \frac{\mathcal I_\alpha^+[\kappa;t]} {\mathcal I_\alpha^{\mathrm{tot}}[\kappa;T]},
\\
p_{\alpha,\kappa}^{-}(t) = \frac{\mathcal I_\alpha^-[\kappa;t,T]} {\mathcal I_\alpha^{\mathrm{tot}}[\kappa;T]} = 1-p_{\alpha,\kappa}^{+}(t).
\label{eq:normalized_past_future_weights}
\end{gathered}
\end{equation}
We combine them into the normalized past-future mixing
\begin{equation}
\mu_{\alpha,\kappa}(t) = 4p_{\alpha,\kappa}^{+}(t)p_{\alpha,\kappa}^{-}(t) = \frac{4\mathcal I_\alpha^+[\kappa;t] \mathcal I_\alpha^-[\kappa;t,T]}{\left[\mathcal I_\alpha^{\mathrm{tot}}[\kappa;T]\right]^2}.
\label{eq:past_future_mixing}
\end{equation}
The factor of four makes $\mu_{\alpha,\kappa}=1$ when the elapsed and remaining contributions are equal. The mixing becomes small when almost all of the coupling lies on one side of $t$. It therefore shows, on a common scale, how sharply a mode is localized between its arrival and departure. 

The mode-selection criteria themselves use the unnormalized coupling weights; $\mu$ is used only for visualization and comparison.

\begin{figure*}
    \centering
    \includegraphics[width=0.98\textwidth]{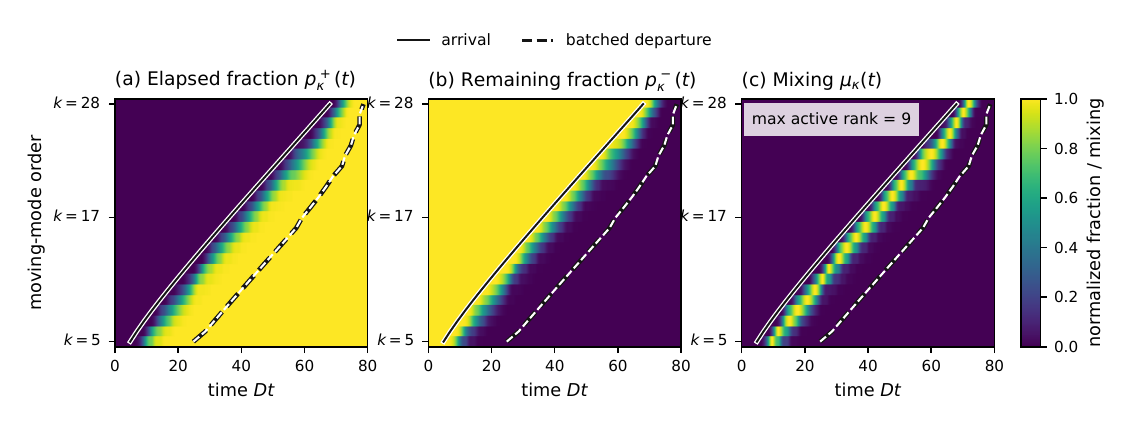}
    \caption{
        Two-sided temporal localization in the corrected tape-recorder moving frame. Panels (a)-(c) show $p_{\alpha,\kappa}^{+}(t)$, $p_{\alpha,\kappa}^{-}(t)$, and $\mu_{\alpha,\kappa}(t)$ for the same 24 completed bulk modes, ordered by increasing equal-weight time. Solid curves mark arrival times obtained from the unnormalized elapsed-weight criterion. Dashed curves mark conservative batched departures, each triggered by a raw crossing of the unnormalized remaining-weight criterion. The same $r_{\mathrm{cut}}=10^{-6}$ is used for arrival and raw-departure selection. The normalized fields are used only for visualization. The corrected raw and batched maximum active ranks are both nine.
    }
    \label{fig:tape_elapsed_remaining_mixing}
\end{figure*}

Figure~\ref{fig:tape_elapsed_remaining_mixing} shows these quantities for the moving-frame modes. The solid curves mark the arrival times obtained from the unnormalized elapsed-weight criterion. The dashed curves mark the conservative batched departure times used by the finite-duration moving frame. Each departure batch is triggered by a raw crossing of the unnormalized remaining-weight criterion. Before arrival the mode is incoming, between the two boundaries it is active, and after departure it is outgoing.

\begin{figure*}
    \centering
    \includegraphics[width=0.8\textwidth]{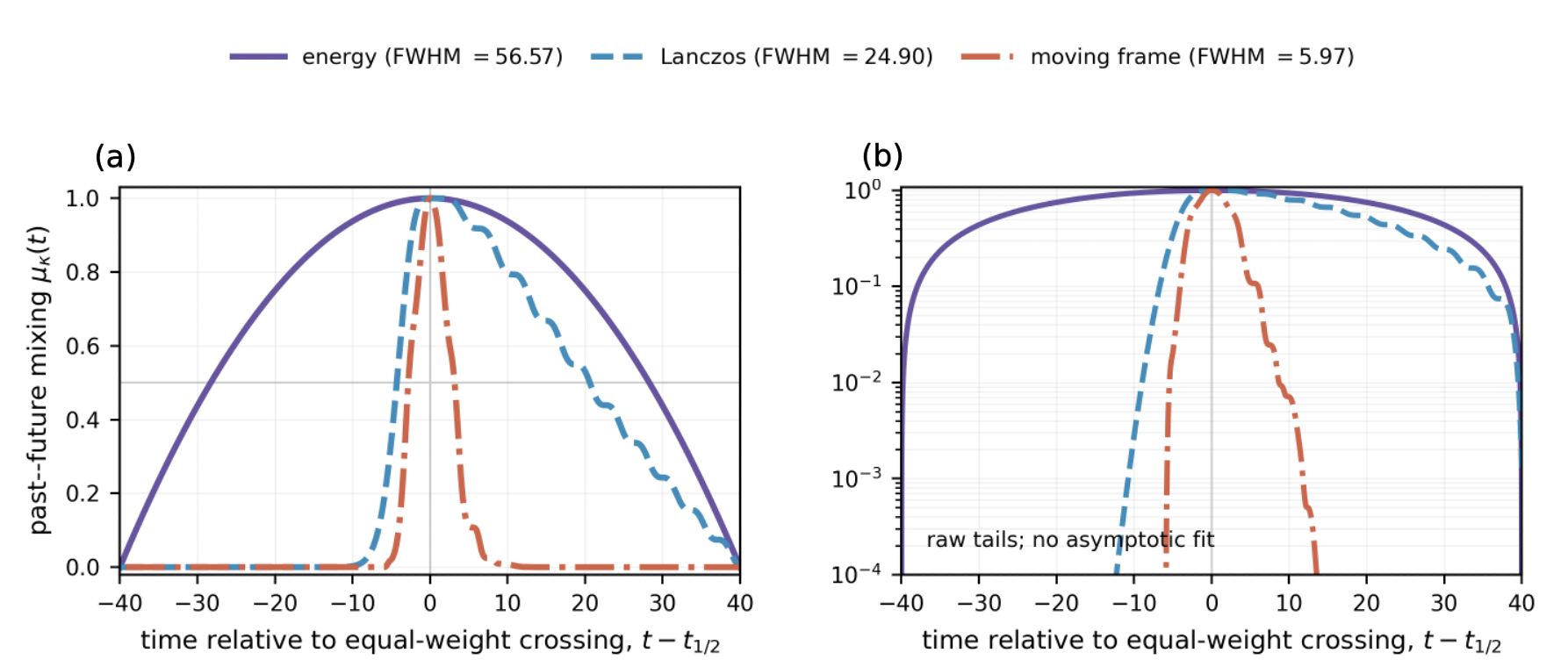}
    \caption{
        Past-future mixing of representative bulk modes in the energy, Lanczos, and moving-frame bases. Time is centered separately at each mode's equal-weight crossing $t_{1/2}$. (a) $\mu_\kappa(t)$ on a linear scale, with the FWHM reported in the legend. (b) The same raw curves on a logarithmic scale, without a tail fit. The moving-frame packet has a much narrower central crossover and a sharper future edge than the Lanczos site. The equal-weight crossing aligns the full coupling histories and is distinct from the algorithmic arrival and departure times. Parameters are the same as in Fig.~\ref{fig:temporal_localization_bases}.
    }
    \label{fig:representative_mode_mixing}
\end{figure*}

The width of $\mu$ provides a common comparison of reservoir representations. As shown in Fig.~\ref{fig:temporal_localization_bases}, the moving frame reduces the bulk median FWHM by factors of about $11.2$ and $4.3$ relative to the energy and Lanczos chain representations, respectively. To examine the shape and tails of an individual bulk mode, we compare one representative mode from each basis in Fig.~\ref{fig:representative_mode_mixing}. The FWHM is $56.57$, $24.90$, and $5.97$ for the energy, Lanczos, and moving-frame modes, respectively. The Lanczos mode retains a long future tail, whereas the moving-frame mode has a much sharper outgoing edge. The logarithmic panel shows the raw tails without assuming a particular decay law.

\section{Forward light cone saturation}\label{sec:saturation}
\begin{statement}
We define the forward light-cone density matrix as
\begin{equation}
    \rho_+(T)=\int_0^T dt\, |\alpha(t)\rangle\langle \alpha(t)|.
\end{equation}

We assume that the bath correlation function

\begin{equation}
    C(t-s)=\langle \alpha(t)|\alpha(s)\rangle
\end{equation}
has spectral representation

\begin{equation}
    C(\tau)=\int_{-\infty}^\infty d\omega\, J(\omega)e^{-i\omega \tau}.
\end{equation}

If the coupling spectral density is essentially bounded,

\begin{equation}
    J\in L^\infty(\mathbb R),
\end{equation}
then the largest eigenvalue of $\rho_+(T)$ is bounded by:
\begin{equation}
    \lambda_{\max}\bigl(\rho_+(T)\bigr)
    \le
    2\pi \|J\|_\infty
\end{equation}
for all \(T\). In particular, \(\lambda_{\max}(\rho_+(T))\) converges to a finite limit as \(T\to\infty\).
\end{statement}

\emph{Proof.} Define the operator

\begin{equation}
    A_T:L^2([0,T])\to \mathcal H_{\mathrm B},
\end{equation}
where $H_{\mathrm B}$ is the bath Hamiltonan, by

\begin{equation}
A_T f = \int_0^T dt\, f(t)|\alpha(t)\rangle .
\end{equation}

Then
\begin{equation}
    \rho_+(T)=A_TA_T^\dagger .
\end{equation}
The nonzero eigenvalues of \(A_TA_T^\dagger\) coincide with those of
\(A_T^\dagger A_T\). The latter is the integral operator
\begin{equation}
(K_T f)(t) = \int_0^T ds\, C(t-s)f(s), \quad t\in[0,T].
\end{equation}
Therefore
\begin{equation}
\lambda_{\max}\bigl(\rho_+(T)\bigr) = \|K_T\|_{L^2([0,T])\to L^2([0,T])}.
\end{equation}

Now let \(f\in L^2([0,T])\), and extend it by zero outside the interval
\([0,T]\). Using the spectral representation of \(C\), we obtain
\begin{align}
    \langle f,K_Tf\rangle
    &=
    \int_0^T dt\int_0^T ds\,
    \overline{f(t)}\,C(t-s)\,f(s)
    \\
    &=
    \int_{-\infty}^\infty d\omega\, J(\omega)
    \left|
        \int_0^T ds\, f(s)e^{i\omega s}
    \right|^2 .
\end{align}
Since \(J(\omega)\le \|J\|_\infty\) almost everywhere,
\begin{equation}
\langle f,K_Tf\rangle \le \|J\|_\infty \int_{-\infty}^\infty d\omega\, \left| \int_0^T ds\, f(s)e^{i\omega s} \right|^2 .
\end{equation}
By Plancherel's theorem:
\begin{equation}
\begin{gathered}
\int_{-\infty}^\infty d\omega\,
\left|\int_0^T ds\, f(s)e^{i\omega s} \right|^2 = \\
= 2\pi \int_{-\infty}^\infty ds\, |f(s)|^2  = 2\pi \int_0^T ds\, |f(s)|^2 .
\end{gathered}
\end{equation}
Hence
\begin{equation}
\langle f,K_Tf\rangle \le 2\pi \|J\|_\infty \|f\|_2^2 .
\end{equation}
Since \(K_T\) is positive,
\begin{equation}
\|K_T\| = \sup_{\|f\|_2=1}\langle f,K_Tf\rangle \le 2\pi \|J\|_\infty .
\end{equation}
Thus
\begin{equation}
\lambda_{\max}\bigl(\rho_+(T)\bigr) \le 2\pi \|J\|_\infty ,
\end{equation}
uniformly in \(T\).

It remains to show convergence. For \(T_2>T_1\),
\begin{equation}
\rho_+(T_2)-\rho_+(T_1) = \int_{T_1}^{T_2} dt\, |\alpha(t)\rangle\langle \alpha(t)| \ge 0 .
\end{equation}
Therefore \(\lambda_{\max}(\rho_+(T))\) is monotone nondecreasing in \(T\). Since it is also bounded above by \(2\pi\|J\|_\infty\), it has a finite limit:
\begin{equation}
\lambda_{\max}\bigl(\rho_+(T)\bigr) \nearrow \lambda_\infty \le 2\pi \|J\|_\infty .
\label{jjj}
\end{equation}

\section{Derivation of the single-event full-state bound}
\label{sec:error_bounds}

This appendix derives Eq.~\eqref{eq:infidelity_bound} for the outgoing-mode truncation defined in Sec.~\ref{sec:tape_recorder}. The exact and truncated states start from the common state $|\Psi_d\rangle$ at the departure time $t_d$ and evolve under $\hat H$ and $\hat H_\perp$, respectively.
Define
\begin{equation}
|\delta\Psi_{\alpha,p}(s)\rangle = |\Psi(s)\rangle-|\Psi_\perp(s)\rangle.
\label{a}
\end{equation}
The exact Duhamel identity gives:
\begin{equation}
\ket{\delta\Psi_{\alpha,p}(s)} = -i\int_{t_d}^{s}d\tau\, \hat U_\perp(s,\tau)\Delta\hat H_{\alpha,p}(\tau)\ket{\Psi(\tau)}
\label{eq1}
\end{equation}
The two terms in Eq.~\eqref{eq:residual_interaction} are placed in the response channels of Eq.~\eqref{eq:finite_response_factors} by defining
\begin{equation}
\begin{aligned}
|F_{\alpha,-}(\tau)\rangle
&= \eta_{\alpha,p}(\tau)
\hat\kappa_{\alpha,p}^{\mathrm{out}} |\Psi(\tau)\rangle,
\\
|F_{\alpha,+}(\tau)\rangle
&= -\eta_{\alpha,p}^{*}(\tau)
\hat\kappa_{\alpha,p}^{\mathrm{out}\dagger} |\Psi(\tau)\rangle .
\end{aligned}
\end{equation}
The minus sign in the second source is produced when the fermionic QPC operator is anticommuted through the outgoing-mode creation operator; it has no effect on the norm.

Let
$\hat N_{\alpha,p}^{\mathrm{out}}
=\hat\kappa_{\alpha,p}^{\mathrm{out}\dagger}
 \hat\kappa_{\alpha,p}^{\mathrm{out}}$.
The squared source norms are
\begin{align}
E_{\alpha,-}(s) 
&\equiv \int_{t_d}^{s}d\tau\, \|F_{\alpha,-}(\tau)\|^2 
\nonumber\\
&= \int_{t_d}^{s}d\tau\,|\eta_{\alpha,p}(\tau)|^2\langle\hat N_{\alpha,p}^{\mathrm{out}}\rangle_\tau,
\\
E_{\alpha,+}(s) &\equiv \int_{t_d}^{s}d\tau\, \|F_{\alpha,+}(\tau)\|^2 
\nonumber\\
&= \int_{t_d}^{s}d\tau\, |\eta_{\alpha,p}(\tau)|^2 \left[1- \langle\hat N_{\alpha,p}^{\mathrm{out}}\rangle_\tau\right].
\label{eq:app_single_event_source_weights}
\end{align}
Here the expectation values are taken in the exact state $|\Psi(\tau)\rangle$. The canonical relation $\hat\kappa\hat\kappa^\dagger=1 - \hat\kappa^\dagger\hat\kappa$ therefore gives:
\begin{equation}
E_{\alpha,-}(s)+E_{\alpha,+}(s) = \int_{t_d}^{s}d\tau\, |\eta_{\alpha,p}(\tau)|^2 = \varepsilon_{\alpha,p}(s).
\label{eq:sess}
\end{equation}
This identity holds for any occupation of the outgoing fermionic mode.

Applying the response-factor bounds to Eq.~\eqref{eq1} and using the triangle inequality gives:
\begin{equation}
\|\delta\Psi_{\alpha,p}(s)\|
\leq C_{\alpha,-}(s)\sqrt{E_{\alpha,-}(s)} + C_{\alpha,+}(s)\sqrt{E_{\alpha,+}(s)}.
\label{eq:app_single_event_two_channel_bound}
\end{equation}

The Cauchy-Schwarz inequality in the two-dimensional channel space and Eq.~\eqref{eq:sess} yield:
\begin{equation}
\|\delta\Psi_{\alpha,p}(s)\|\leq\sqrt{C_{\alpha,-}^2(s)+C_{\alpha,+}^2(s)}\,\sqrt{\varepsilon_{\alpha,p}(s)}.
\label{eq:single_event_state_response_bound}
\end{equation}
For normalized vectors,
$1-|\langle\phi|\psi\rangle| \leq\tfrac12\|\,|\phi\rangle-|\psi\rangle\,\|^2$.
Combining this inequality with Eq.~\eqref{eq:single_event_state_response_bound} proves Eq.~\eqref{eq:infidelity_bound}.

For completeness, unitarity gives an explicit finite-interval estimate for the response factors. Let $\hat X_\alpha$ denote either $\hat V_\alpha^\dagger$ or $\hat V_\alpha$. Since $\|\hat X_\alpha\|=\|\hat V_\alpha\|$,
\begin{equation}
\begin{gathered}
\left\| \int_{t_d}^{s}d\tau\,\hat U_\perp(s,\tau)\hat X_\alpha F(\tau)\right\|\leq\|\hat V_\alpha\|\int_{t_d}^{s}d\tau\,\|F(\tau)\|
\\
\leq \sqrt{s-t_d}\,\|\hat V_\alpha\|\left(\int_{t_d}^{s}d\tau\,\|F(\tau)\|^2 \right)^{1/2}.
\label{eq:app_explicit_response_bound}
\end{gathered}
\end{equation}

Thus $C_{\alpha,\pm}(s) \leq\sqrt{s-t_d}\,\|\hat V_\alpha\|$, which gives Eq.~\eqref{eq:infidelity_bound_linear}. The departure criterion then yields Eqs.~\eqref{eq:single_event_threshold_response_bound} and~\eqref{eq:single_event_threshold_explicit_bound} at $s=T$.

\bibliography{bibliography}

\end{document}